\documentclass[useAMS,usenatbib]{mn2e} 
\usepackage{graphicx} 
\title[Effect of Debris on Terrestrial Planet Formation]{Planetesimals to Protoplanets II: Effect of Debris on Terrestrial Planet Formation}
\author[Z. M. Leinhardt, D. C. Richardson, G. Lufkin, and J. Haseltine]{Z. M. Leinhardt$^1$, D. C. Richardson$^2$, G. Lufkin$^2$, and J. Haseltine$^2$\\
$^1$ Department of Applied Mathematics and Theoretical Physics, University of Cambridge, Cambridge, CB3 0WA, U.K.\\
$^2$ Department of Astronomy, University of Maryland, College Park, MD 20742, USA\\
}

\begin{document}

\date{Submitted to MNRAS 16 December 2008, accepted 11 March 2009}

\pagerange{\pageref{firstpage}--\pageref{lastpage}} \pubyear{}

\maketitle

\label{firstpage}

\begin{abstract}
In this paper we extend our numerical method for simulating terrestrial planet formation from \citet{Leinhardt05} to include dynamical friction from the unresolved debris component. In the previous work we implemented a rubble pile planetesimal collision model into direct $N$-body simulations of terrestrial planet formation. The new collision model treated both accretion and erosion of planetesimals but did not include dynamical friction from debris particles smaller than the resolution limit for the simulation. By extending our numerical model to include dynamical friction from the unresolved debris, we can simulate the dynamical effect of debris produced during collisions and can also investigate the effect of initial debris mass on terrestrial planet formation. We find that significant initial debris mass, 10\% or more of the total disk mass, changes the mode of planetesimal growth. Specifically, planetesimals in this situation do not go through a runaway growth phase. Instead they grow concurrently, similar to oligarchic growth. The dynamical friction from the unresolved debris damps the eccentricities of the planetesimals, reducing the mean impact speeds and causing all collisions to result in merging with no mass loss. As a result there is no debris production. The mass in debris slowly decreases with time. In addition to including the dynamical friction from the unresolved debris, we have implemented particle tracking as a proxy for monitoring compositional mixing. Although there is much less mixing due to collisions and gravitational scattering when dynamical friction of the background debris is included, there is significant inward migration of the largest protoplanets in the most extreme initial conditions (for which the initial mass in unresolved debris is at least equal to the mass in resolved planetesimals).

\end{abstract}

\section{Introduction}

Extrasolar planets are numerous and diverse, with recently detected examples ranging from ``hot Jupiters", with periods of days, to super-Earths, with masses $\ge 5 M_\oplus$ \citep[][]{Anderson08,Mayor08}. It is evident that planet formation is common and ubiquitous. However, results from numerical simulations lag behind recent observational discoveries. Simulations of the formation of our own solar system often result in planetary eccentricities that are too high and/or ejection of one of the terrestrial planets \citep{Raymond08}.   
Damping of eccentricities requires either gas \citep{Tanaka04} or large amounts of small debris \citep{Goldreich04}. It has yet to be shown quantitatively whether either situation arose in our own solar system. 

Numerical simulations permit modeling the complex interplay of physical processes during planet formation that are otherwise difficult to assess. Observations are snapshots that provide only limited information on how a protoplanetary disk evolves into a solar system. In addition, not all phases of planet formation are observable. Our approach to understanding the formation and evolution of solar systems is to begin with a simple model and systematically make the model more and more realistic. In this paper, as with our previous paper \citep{Leinhardt05}, we have chosen to focus on the terrestrial region during the middle phase of planet formation, assuming no gas but allowing for the production of background debris. Thus in this work we ignore the effects of gas. In this context we use an $N$-body code to model the collisional and dynamical evolution. 

\subsection{Previous Work}

\citet{Kokubo02} completed a series of direct $N$-body simulations of the middle phase of terrestrial planet formation \citep[for reviews see][]{Lissauer93,Chambers04}. In these simulations it was assumed that every collision resulted in perfect merging (no mass loss), thus, every collision resulted in growth of the planetesimals. There was no erosion nor the possibility of producing debris that could damp the larger protoplanets found at the end of the simulations.  In our previous paper we developed a more realistic planetesimal collision model that allowed both accretion and erosion. We then ran evolution models with similar starting conditions to \citet{Kokubo02}. We found virtually identical results.  The number, mass, and spatial distribution of the protoplanets were similar and showed both runaway and oligarchic growth. However, this earlier work did not include a model for the full dynamical feedback of the collisional debris (mass elements below the simulation resolution) on the planetesimals. Though planetesimals could accrete collisional debris, there was no dynamical friction from the debris on the planetesimals. In addition, all simulations in \citet{Leinhardt05} used a low mass of initial debris: the effect of starting with a larger amount of mass as debris was not investigated. 

The middle phase of planet formation is not observable (planetesimals are too big to be observed by infrared observations and too small to be observed at visible wavelengths). In addition, the physical mechanisms that govern the earlier phases, namely planetesimal formation, are disputed \citep[for example, the long-standing debate between gravitational instability and turbulence models;][]{Johansen07, Cuzzi08}. As a result, the initial conditions for numerical simulations of this phase are unknown. In general, it is usually assumed that planetesimals that have just decoupled from the gas are on similar, almost circular orbits with small inclinations. However, the amount of material in planetesimals and the amount in smaller debris is completely unknown, and as with many of the protoplanetary disk parameters, the amount of debris may vary from star to star depending on metallicity and/or spectral type. \citet{Leinhardt05} chose an initially low mass in debris because they did not have full feedback in their numerical model. 

In this paper we investigate the effect of the initial debris distribution on the middle phase of planet formation. We use a similar numerical method as \citet{Leinhardt05} but one that is upgraded to include dynamical friction from unresolved material and the ability to track compositional mixing in both the planetesimals and the debris through the course of the simulation.
 
\section{Method}\label{sec:method}

For our simulations we use the parallel $N$-body gravity code \texttt{pkdgrav}. The code uses a second-order leap-frog integrator with a hierarchical tree \citep{Richardson00,Stadel01} to provide computation time scaling as $O(N\log N)$, where $N$ is the number of resolved planetesimals. The version used here has been modified to include realistic collisions between planetesimals \citep[see][for details]{Leinhardt05} and accounts for dynamical friction from the collisional debris. 

\subsection{Planetesimal Model}\label{sec:model}

In this section we summarize the planetesimal structure and collision model used in the numerical simulations. The planetesimal model used in this paper is the same as that used in \citet{Leinhardt05}. Please refer to that paper for more details.

Due to numerical limitations, our direct numerical simulations must start with large planetesimals ($\sim$ 60 km in radius with bulk density 2 g cm$^{-3}$). As a result, these planetesimals are in the gravity regime where their material strength is negligible compared to their gravitational strength \citep[][and references therein]{Leinhardt08,Leinhardt09,Stewart09}. Thus, we have chosen to model the planetesimals in our numerical simulations as gravitational aggregates \citep[or, more precisely, idealized rubble piles---cf.][]{Richardson02} during collisions.

In the numerical simulations presented here, planetesimals grow via accretion of debris and  planetesimal--planetesimal collisions. The collisions between planetesimals are compared to a look-up table of collision outcomes, and either the tabulated result is used (if suitable), or a direct simulation of the interaction is performed. Collision parameters of impact speed, impact parameter, and mass ratio, as well as a user-specified coefficient of restitution are used to interpolate or extrapolate the collision outcome from the collision outcome database. The database contains the mass of the largest post-collision remnant from several hundred rubble-pile planetesimal collision simulations over a wide range of parameter space \citep{Leinhardt00,Leinhardt02,Leinhardt05}.

If the predicted collision outcome from the database is one large remnant and a small amount of debris, the colliding particles are replaced by the large remnant and the debris is followed semi-analytically (see below). If, on the other hand, the collision results in two massive bodies (the largest post-collision remnant being less than 80\% and second largest remnant being greater than 20\% of the total mass of the system), the planetesimals---which have been modeled as single particles up to this point---are replaced by rubble piles of 100 or so particles each, and the collision is integrated explicitly in the planetesimal disk. 

In this latter case of a resolved collision, for the first ten dynamical times ($T_\mathrm{dyn} \simeq 1/\sqrt{G \rho}$, where $G$ is the gravitational constant and $\rho$ is the bulk density of the planetesimals), the particles are only allowed to bounce inelastically (there is no merging, in order to allow the system to relax in a realistic way, nor fracturing, because the re-impact speeds between particles are small). After ten dynamical times, the rubble particles merge with one another when they collide. After twenty dynamical times, any particle from the collision that is smaller than the resolution limit (the size of a planetesimal at the start of the simulation) is demoted to unresolved debris. The time allotted for these bouncing and merging phases is based on studying the collision experiments used to generate the outcome database. It was found that in most cases intervals of ten dynamical times were an adequate compromise between detailed collision outcome modeling and computational expediency.

In our model, the planetesimal disk is divided into a configurable number of annuli for the purpose of tracking debris. Unresolved debris from a collision is added to the mass density of the annulus at the location where the debris was generated. The debris is assumed to have circular Keplerian orbits and a fixed scale height of $1\times 10^{-4}$ AU. The value of the scale height was chosen such that all coplanar planetesimals are fully embedded in the debris disk during the entire evolution of the system. The planetesimals sweep up debris as they pass through the annuli  \citep[see \S 2.3 of][]{Leinhardt05} and are damped by the dynamical friction of the debris (\S \ref{sec:dynfric} below). If a collision occurs at the edge of the debris simulation domain (either interior to the inner edge of the inner annulus or exterior to the outer edge of the outer annulus) any unresolved debris outside the debris simulation domain after twenty dynamical times is put into a ``trash bin" for tracking purposes. The mass density and composition of the trash bin are tracked through the entire simulation like any other annulus bin but the contents do not interact with the planetesimals. The resolved planetesimals are free to move inside or outside of the debris simulation domain. If they move outside, the resolved planetesimals will not interact with any debris during that interval.

\subsection{Dynamical Friction}\label{sec:dynfric}

Unresolved debris circularizes the planetesimals through the action of periodic impulses via equation 7-14 of \citet[][]{Binney87},
\begin{equation}\label{eqn:dynfric}
\frac{d\bf{v}_M}{dt} = 8\pi^2\ln (1+ \Lambda^2) G^2m(M+m)\frac{\int_0^{v_M} f(v_m)v_m^2\, dv_m}{v_M^3}\bf{v_M}
\end{equation}
where the $\bf{v}_M$ is the difference between the orbital velocity of the resolved planetesimal and that of the unresolved debris at the planetesimal's instantaneous position, $m$ is the mass of a single unresolved debris particle, $M$ is the mass of the planetesimal, and the integral gives the amount of unresolved material with speed less than the speed of the planetesimal, for an assumed debris particle speed distribution $f(v_m)$. In the Coulomb logarithm for a Keplerian disk, $\Lambda = b_{max}(v_M^2 + 2 V_{disp} v_{circ}^2)/(GM)$, the maximum impact parameter between a planetesimal and the unresolved debris is $b_{max} = R_H + a \sqrt{V_{disp} + \sin^2 i}$, where $R_H = (\frac{M}{3 M_\odot})^{1/3}$ is the Hill radius of the planetesimal, $a$ is the semi-major axis of the planetesimal, $i$ is the inclination of the planetesimal, $V_{disp}$ is the velocity dispersion of the unresolved debris, $v_{circ}$ is the instantaneous circular speed of the planetesimal, and $G$ is the gravitational constant \citep{Stewart00,Ford07}. 

In this paper we assume the unresolved debris is ``cold"  (having zero eccentricity and inclination). Thus, the relative speed of the planetesimal to the mean Keplerian speed of the debris disk is generally much larger than the velocity dispersion of the unresolved debris. Let us assume that the unresolved debris has a Maxwellian distribution \citep[Eq.~7-16][]{Binney87},
\begin{equation}
f(v_m) = \frac{n_0}{(2 \pi V_{disp}^2)^{3/2}} \exp{(-\frac{1}{2}v_m^2/V_{disp}^2)},
\end{equation} 
where $n_0$ is the number density of unresolved debris particles.
We may then integrate Eq.~\ref{eqn:dynfric} \citep[see equation 7-17][]{Binney87},
\begin{equation}\label{eqn:inter}
\frac{d\bf{v}_M}{dt} = \frac{-2\pi \ln(1 + \Lambda^2) G^2 \rho M}{v_M^3}[ \rm{erf}(X) - \frac{2 X}{\sqrt{\pi}} e^{-X^2}]{\bf v_M},
\end{equation}
where $X = v_M/(\sqrt{2} V_{disp})$, $\rho = n_0 m $ is the mass density of the unresolved debris particles, and $m \ll M$, therefore, $M + m \sim M$. We now apply the assumption that the $V_{disp} \ll v_M$ (see \S \ref{sec:disc} for discussion of the relaxation of this condition), so $X$ goes to infinity and Eq.~\ref{eqn:inter} becomes,
\begin{equation}\label{eqn:dynfricII}
\frac{d\bf{v}_M}{dt} = \frac{-2\pi \ln(1 + \Lambda^2) G^2 \rho M}{v_M^3}{\bf v_M}.
\end{equation}

In the simulations presented in this paper we have sped up the planetesimal growth by artificially inflating the radii of the planetesimals by an expansion parameter $f$ \citep[][]{Kokubo02,Leinhardt05}. As a result, the dynamical friction (Eq.~\ref{eqn:dynfricII}) must be modified as well in order to keep pace with the accelerated collisional evolution. The collision cross-section scales roughly as the surface area for most of the evolution modeled in these simulations until gravitational focusing becomes important. Thus, the dynamical friction impulse applied becomes
\begin{equation}\label{eqn:dynfricIII}
\Delta{\bf v}_M = \frac{-2\pi \ln(1 + \Lambda^2) \rho G^2M}{v_M^3}f^2\, \Delta t \, {\bf v_M}
\end{equation}
where $\Delta t$ is the time between impulse applications. In all simulations presented here, $f=6$. $\Delta t$ varies between 1 and 10 timesteps depending on the initial mass of the debris and the magnitude of the resulting impulse (see \S \ref{sec:time} \& Table \ref{tab:sims} for more details about time steps).

\subsection{Composition Tracking}\label{sec:origin}

In addition to the implementation of full dynamical friction, we have added the capability to track mixing both in planetesimals and the unresolved debris. Each planetesimal carries a mass-weighted fractional composition histogram that is binned by the semi-major axis. The number of histogram bins is user-specified---in all simulations presented here we used 15 bins. Each annulus of unresolved debris carries an analogous composition histogram. At time zero, all bins in each histogram have zero value except for the bin corresponding to the initial particle/debris location, which has a value of one. As a planetesimal grows, the populations of its histogram bins vary according to the evolving composition. For debris, the histogram bins evolve as new debris enters into the annulus, or existing debris is swept up by planetesimals.

\begin{figure*}
 \includegraphics[scale=0.5]{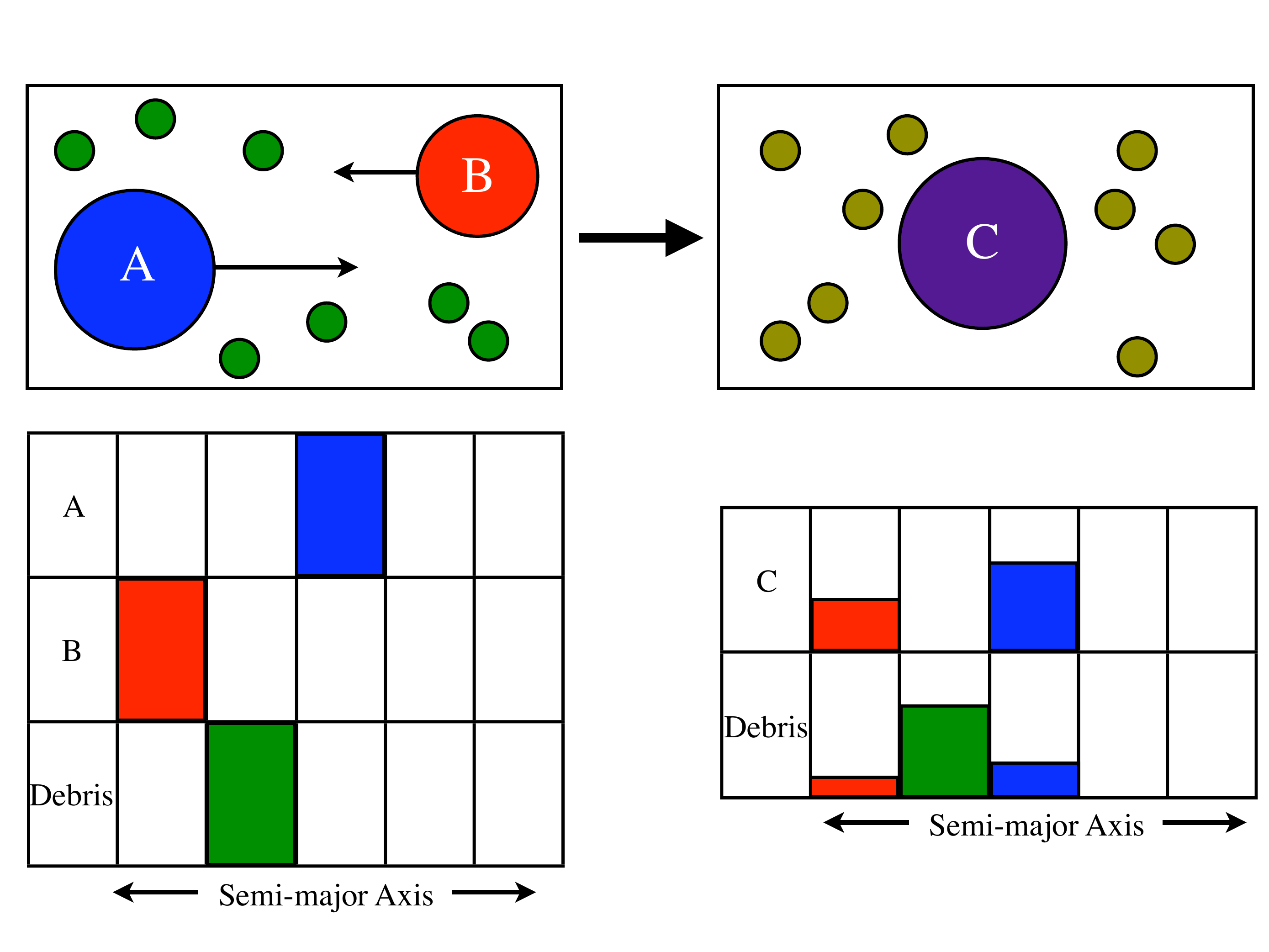}
 \caption{Cartoon of a collision between two planetesimals (top left), A (blue) and B (red), in a swarm of unresolved debris (smaller green circles). The result (top right) is planetesimal C (purple) and heterogeneous debris (small olive-color circles). Example composition histograms pre- and post-collision are shown on the bottom left and right, respectively. 
 \label{fig:origin}}
 \end{figure*}

Figure \ref{fig:origin} shows a cartoon of a collision between two planetesimals (A \& B) in a swarm of background debris (small unlabeled circles). The composition histograms for the planetesimals and debris are shown below the cartoon of the collision. In the example, there are five semi-major axis bins. Planetesimal A is originally made up completely of material from the middle of the protoplanetary disk (blue rectangle in lower left of Fig.~\ref{fig:origin}). Planetesimal B is originally made up of material from the inner annulus (red rectangle), and the debris is originally made up entirely of material in between A and B (green rectangle). 

After the collision, there is one planetesimal, C, and some additional unresolved debris. Planetesimal C is a mass-weighted compositional mix of planetesimals A \& B. The composition histogram for C (lower right) has some material from planetesimal A (blue rectangle in the C row) and some from planetesimal B (red rectangle in the C row). The debris is also mixed as a result of the collision, now containing material from both original planetesimals (blue and red rectangles) in addition to the original debris material composition. 

A planetesimal that passes through the now heterogeneous mixed debris will sweep some of it up. The composition of the debris will modify the composition histogram of the growing planetesimal, just as if it had been modified in a collision with another planetesimal.

\subsection{Planetesimal Disk Model}

In this paper we present moderate-resolution (initial number of resolved planetesimals $N = 10^4$) simulations of various initial debris disk masses to investigate the effect of environment on protoplanet formation (see \S \ref{sec:general}). For all simulations we used the standard model for the resolved planetesimal disk from \citet{Leinhardt05}. If all of the mass is in resolved planetesimals ($M_r$) and no mass is in unresolved debris ($M_d$), the standard model reduces to a ``minimum-mass solar nebula". For all simulations, the initial surface density of planetesimals at 1 AU,  $\Sigma_1 \sim 10.0$ g cm$^{-2}$, and the surface density distribution $\Sigma_p = \Sigma_1 (a/1\, \rm{AU})^{-3/2}$. The total mass of unresolved debris varied from $0 - 25\, M_\oplus$, with surface density at 1 AU ($\Sigma'_1$) between 0 and 100 g cm$^{-2}$, and a surface density distribution set to $\Sigma_{debris} = \Sigma'_1 (a/1\,  \rm{AU})^{-3/2}$. Each simulation began with a 1 AU-wide band of particles and unresolved debris centered at 1 AU. The simulations were run for at least $1 \times 10^5$ yr---long enough to initiate runaway growth and see the formation of multiple protoplanets in previous work \citep{Leinhardt05,Kokubo02}. 
 
\subsection{Time Step}\label{sec:time}

In the numerical simulations presented here there are two time scales: the orbital time scale of the planetesimals (planetesimal-Sun interactions) and the time scale of planetesimal-planetesimal interactions. The orbital dynamical time is one year at 1 AU. The planetesimal-planetesimal interaction time scale is significantly less ($\sim 1$ hr for $\rho = 2$ g cm$^{-3}$) as it depends inversely on the bulk density of the colliding planetesimals (\S \ref{sec:model}). The minimum step size is a function of the choice of integrator and the dynamical time. For leapfrog integrations of low-eccentricity orbits, $\sim 33$ steps per orbit are required to accurately resolve the orbit \citep{Quinn97}. In \citet{Leinhardt05}, we were more conservative and used $\sim 100$ steps per dynamical time to resolve orbital or collisional interactions between planetesimals. The large difference in the time step required to model collisions and orbits has led us to implement a bi-modal particle time stepping scheme. The large step is used to integrate the orbits of the planetesimals while the small steps are used during the interval between when a collision is detected \citep[for details see \S 2.5 of][]{Leinhardt05} and finally resolved. In this paper we have found that the addition of dynamical friction reduces the minimum time step required to accurately integrate the planetesimal orbits. In other words, $100$ steps per orbit at 1 AU is not sufficient resolution for all initial conditions considered here (discussed below). This is because the impulse (Eq.~\ref{eqn:dynfricIII}) from the dynamical friction of the debris can be large in comparison to the speed of a given planetesimal, resulting in a large change in direction and/or magnitude of the planesimal's velocity vector. The step size required is strongly dependent on the magnitude of the dynamical friction impulse and thus the debris mass.

\begin{figure}
 \includegraphics[scale=0.4]{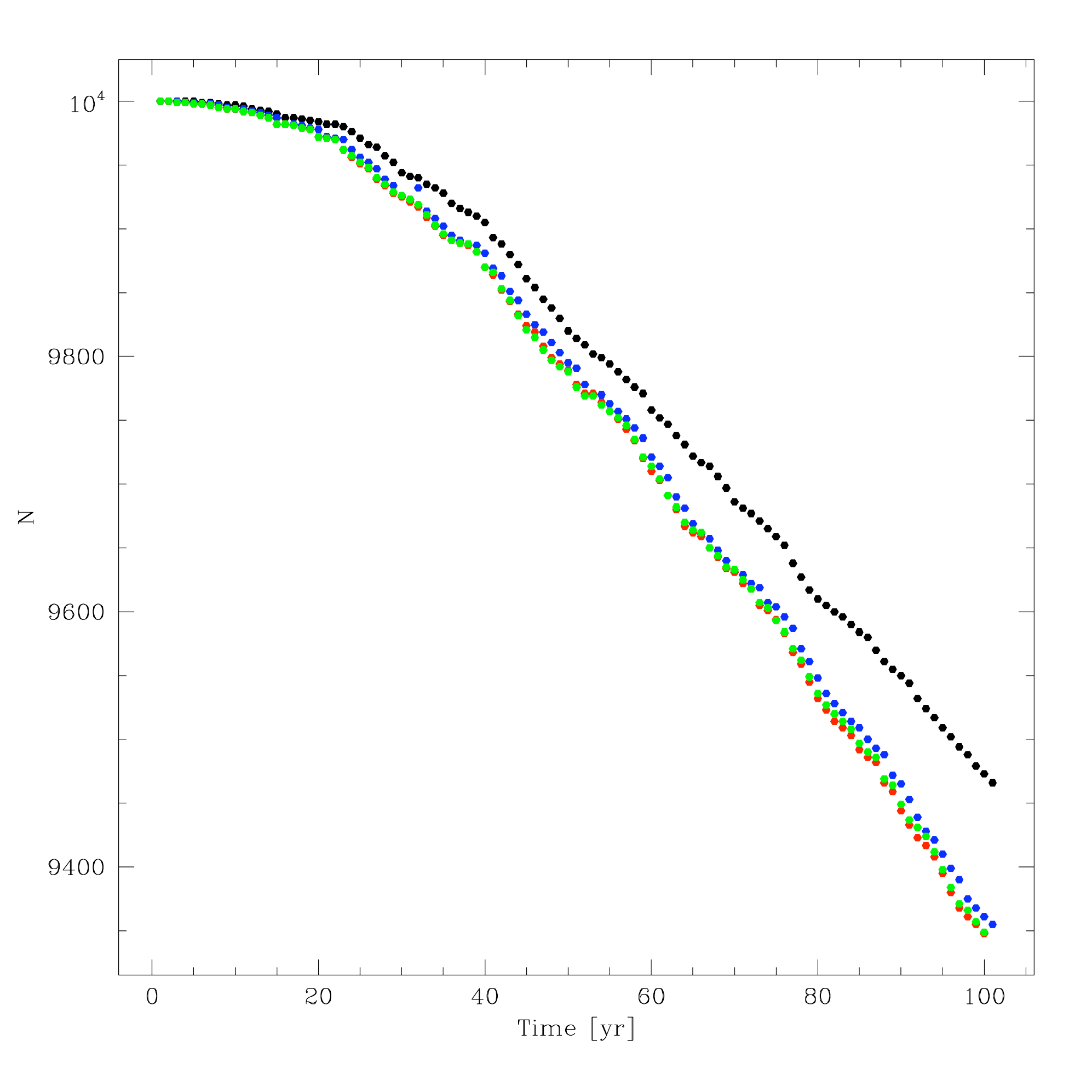}
 \caption{Number of particles versus time using four different time steps. Black: $\Delta t = 2.5\times 10^{-4}$ yr, blue: $\Delta t = 1\times 10^{-4}$ yr, red: $\Delta t = 0.5\times 10^{-4}$ yr, green: $\Delta t = 0.25\times 10^{-4}$ yr. The curve for the simulation with the longest time step gradually separates from those using the smaller time steps, indicating that $2.5 \times 10^{-4}$ yr is too large a step to accurately model the dynamical evolution of the protoplanetary system.  
 \label{fig:time}}
 \end{figure}

For each set of initial conditions we completed a time step test that consisted of comparing the number of particles as a function of time for simulations of the same initial condition but different major (i.e.~orbital) time steps. The time step necessary for a particular initial condition was determined when the number of particles versus time agreed with a simulation of smaller time step. Figure \ref{fig:time} shows the number of particles versus time for a simulation with the initial mass of the unresolved debris equal to 10\% of the mass of planetesimals. The black, blue, red, and green points correspond to major steps of 2.5, 1, 0.5, and $0.25 \times 10^{-4}$ yr, respectively. The evolution of the number of particles in the largest step size simulation is slower than for the smaller step sizes. However, all three smaller step sizes have a similar slope, thus for this simulation a time step of $1 \times 10^{-4}$ yr was used for the major step (see Table \ref{tab:sims} for time steps used in each simulation). 

In addition to the time step test, the angular momentum and energy conservation was checked for a perfect merging simulation that does not include the dissipative effects of dynamical friction from unresolved debris. The fractional angular momentum after $1\times 10^5$ yr was conserved to one part in $10^5$ and the energy to one part in $10^4$. When unresolved debris is included, the angular momentum of the resolved planetesimals increases as the planetesimals accrete unresolved debris, as expected.

\section{Results}

Table \ref{tab:sims} summarizes the chosen parameters of individual simulations. In the table, the first column gives the simulation name while the second column indicates whether dynamical friction from the unresolved debris was included or not ([Y]es or [N]o). The rest of the column labels are as follows: $\Sigma'_1$ is the surface mass density of the unresolved debris at 1 AU, $\Delta t$ is the major time step, $\delta t$ is the minor time step, and $T_\mathrm{tot}$ is the total simulation time. The control simulations (sims.~1-4a) did not include dynamical friction from the unresolved debris and were run using the same numerical method as \citet{Leinhardt05}. Comparisons between the control simulations and those with dynamical friction (1-4b) are presented in \S \ref{sec:general}-\ref{sec:collout}. In \S \ref{sec:general} we present a summary of the changes in planetesimal evolution that occur when dynamical friction of the unresolved debris is included in the model. In \S \ref{sec:mix} we discuss the differences in the composition of the protoplanets after $10^5$ yr. The evolution of the unresolved debris is presented in \S \ref{sec:debris}. In \S \ref{sec:growth} we show that significant initial debris mass changes the growth mode of the planetesimals when the dynamical friction of the unresolved debris is included. In the last section (\S \ref{sec:collout}) we discuss the differences in the collision outcomes between the control simulations and those that include dynamical friction of the unresolved debris.

\begin{table}
\centering
\begin{minipage}{140mm}
\caption{Summary of simulation parameters. \label{tab:sims}}
\begin{tabular}{cccccc}
\hline
Sim. & Dyn. Fric. & $\Sigma'_1$ & $\Delta t$ (yr) & $\delta t$ (yr) & $T_{tot}$ (yr) \\
\hline
1a & N & 0.00 & $1\times 10^{-3}$ & $1.25\times 10^{-4}$ & $2.1\times 10^5$\\
1b & Y & 0.00 & $1\times 10^{-4}$ & $5.00 \times 10^{-5}$ & $1.0 \times 10^5$\\ 
2a & N & 1.00 & $1\times 10^{-3}$ & $1.25\times 10^{-4}$ & $4.0 \times 10^5$\\
2b & Y & 1.00 & $1\times 10^{-4}$ & $5.00 \times 10^{-5}$ & $1.0 \times 10^5$\\
3a & N & 10.0 & $1\times 10^{-4 }$ & $5.00 \times 10^{-5}$ & $1.0 \times 10^5$\\
3b & Y & 10.0 & $1\times 10^{-4}$ & $5.00 \times 10^{-5}$ & $1.0 \times 10^5$\\
4a & N & 100. & $1\times 10^{-4 }$ & $5.00 \times 10^{-5}$ & $1.0 \times 10^5$\\
4b & Y & 100. & $1\times 10^{-4}$ & $5.00 \times 10^{-5}$ & $1.0 \times 10^5$\\

\hline
\end{tabular}
\end{minipage}
\end{table}

\subsection{Effect of Dynamical Friction}\label{sec:general}

\begin{figure*}
 \includegraphics[scale=0.7]{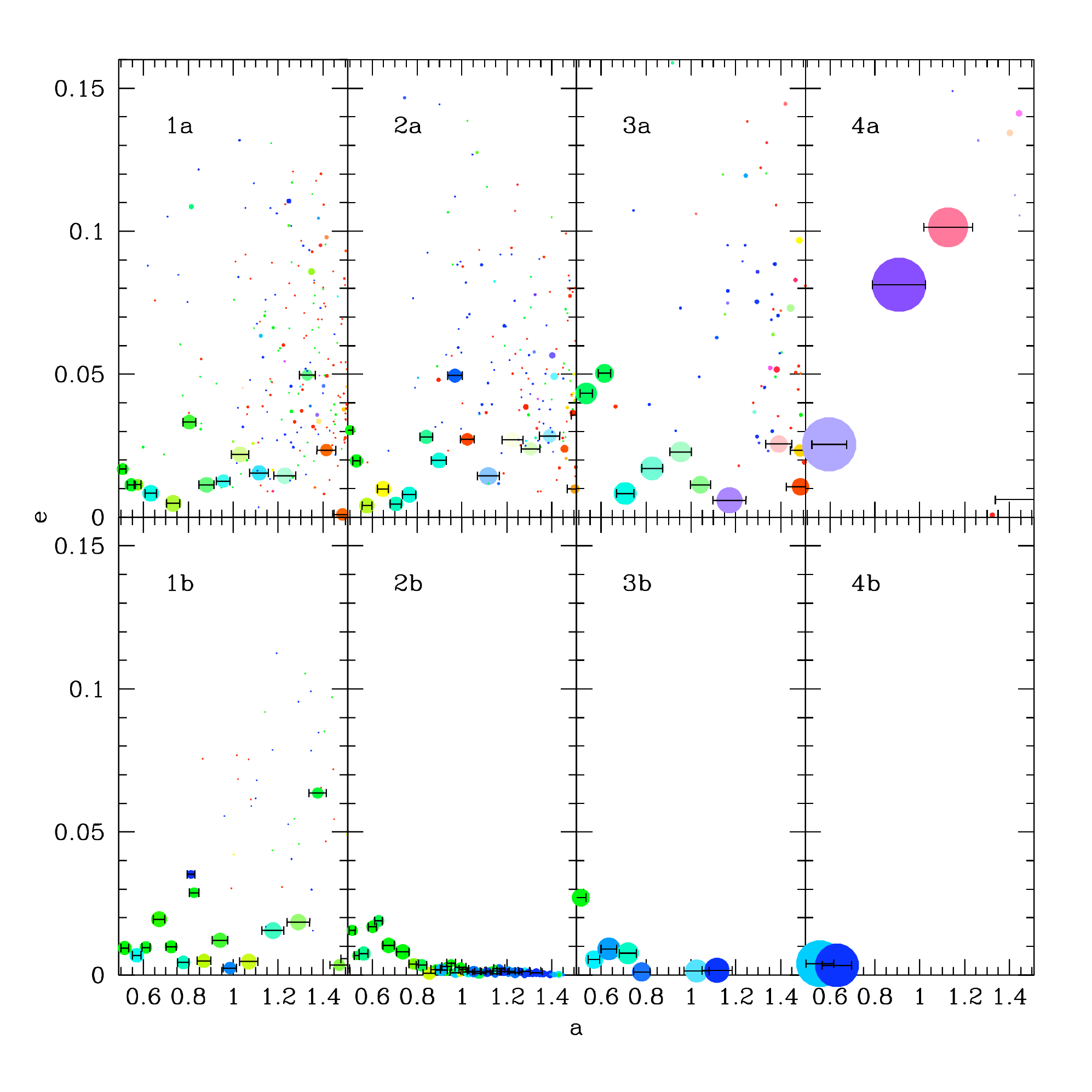}
 \caption{Semi-major axis versus eccentricity of particles after 100,000 yr. The top row shows simulations 1a, 2a, 3a, and 4a. The bottom row shows simulations 1b, 2b, 3b, and 4b. The circles with error bars are particles that have reached masses greater than 100 times the starting planetesimal mass ($m_{\rm o} = 1.5 \times 10^{24}$ g). The error bars are $10 R_\mathrm{H}$ in length, where $R_\mathrm{H} = (2 M/3M_*)^{1/3}a$ is the mutual Hill radius, $M$ is the mass of the protoplanet and $M_*$ is the mass of the star. The colors indicate composition with respect to the particle's current location. Green particles are made of material predominantly from near the particles' current location, blue particles contain material from larger semi-major axis, and red particles contain material from smaller semi-major axis. Non-primary-colored particles represent a mixture (mass-weighted) of material from these three categories. 
 \label{fig:ae}}
 \end{figure*}

\begin{figure}
 \includegraphics[scale=0.43]{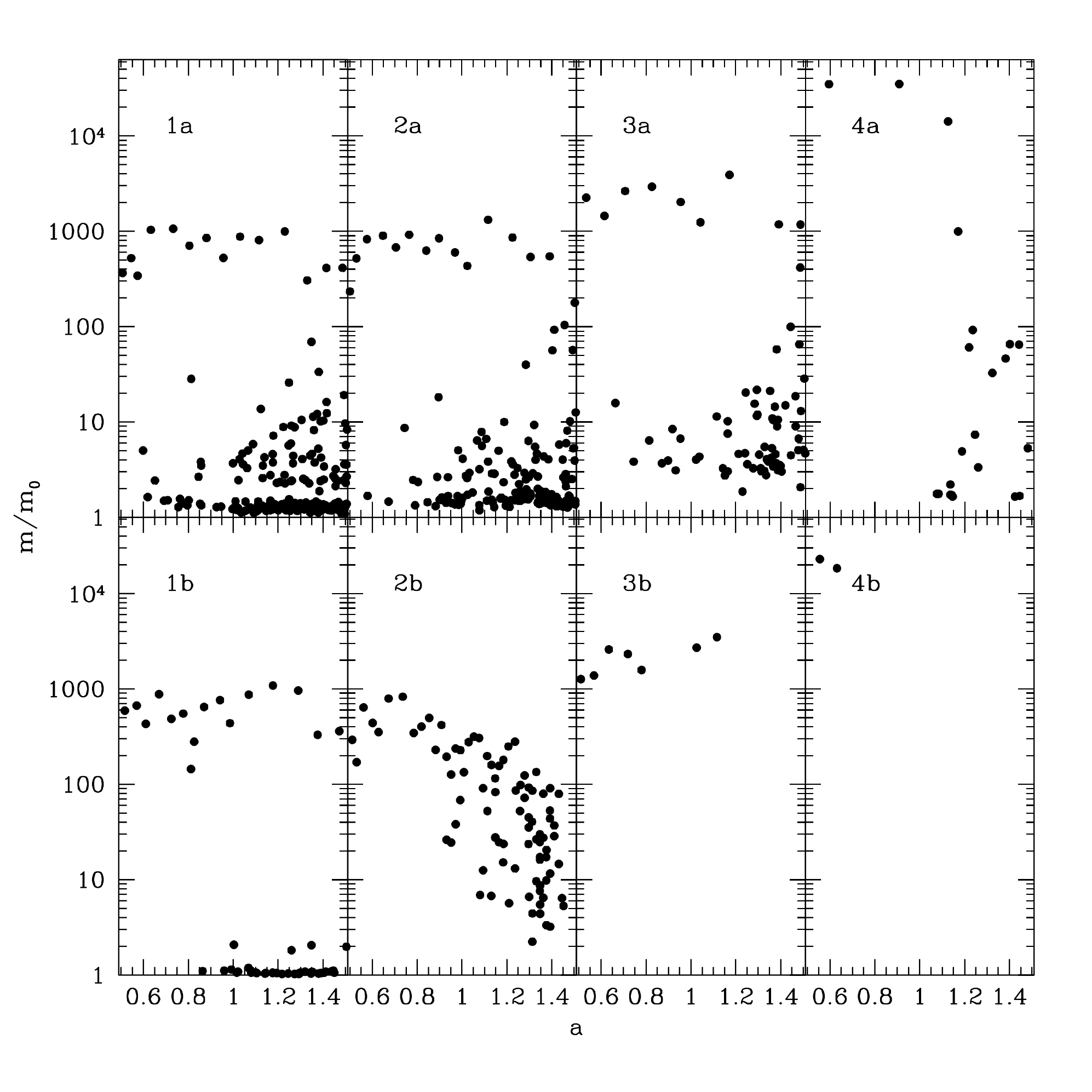}
 \caption{Semi-major axis versus particle mass after 100,000 yr in units of the initial mass of the planetesimals ($1.5 \times 10^{24}$ g). Top panels correspond to simulations 1a, 2a, 3a, and 4a, bottom panels correspond to simulations 1b, 2b, 3b, and 4b.
 \label{fig:am}}
 \end{figure}

Dynamical friction from unresolved debris changes the evolutionary growth of protoplanets by damping eccentricities and, in some cases, causing significant inward migration. Figures \ref{fig:ae} and \ref{fig:am} show the eccentricity and mass of the planetesimals versus semi-major axis after 100,000 yr of simulation. The columns of plots from left to right have initial total mass in unresolved debris of 0 (sims.~1a \& 1b), 0.25 (2a \& 2b), 2.5 (3a \& 3b), and 25 $M_\oplus$ (4a \& 4b). The top row shows the control simulations (1-4a) that do not include dynamical friction from the unresolved debris. The bottom row shows simulations (1-4b) that included dynamical friction. 

The dynamical friction of the background debris significantly damps the eccentricities and causes inward migration of protoplanets (the largest planetesimals; see \S \ref{sec:mix}). The maximum eccentricity of the protoplanets in simulations 2b and 3b is a few times less than the maximum eccentricity of the protoplanets in 2a and 3a (Fig.~\ref{fig:ae}); they are not zero due to a few stirred protoplanets at small semi-major axis. Most of the protoplanets in 2b and 3b and both protoplanets in 4b have effectively zero eccentricity. In addition, the dynamical friction has an equalizing effect on the planetesimals, causing them to evolve (grow) as a single population. There is little or no identifiable background population of small bodies in simulations 1, 2, and 3b compared to a large background in sims 1, 2, and 3a. This is especially noticeable in simulation 2b, which shows a smooth continuum in mass from protoplanets almost $1\times 10^3$ times the starting mass, $m_0$, to planetesimals of mass 2 $m_0$. In the comparison, control simulation 2a shows a clear separation between the protoplanets ($m \sim 1\times 10^3$ $m_0$) and the background planetesimals ($m = 1 - 10$ $m_0$). The protoplanets in simulation 4a have grown large enough that they have agglomerated with or scattered all planetesimals within $10 R_{\mathrm H}$. Simulation 4b is quite different: the two large protoplanets have evidently migrated inward, sweeping up any particles they encounter and leaving no particles with semi-major axis larger than 0.65 AU.

 \subsection{Mixing}\label{sec:mix}
 
After $10^5$ yr the protoplanets in the control simulations are relatively heterogenous in composition, however, when dynamical friction of the unresolved debris is included, mixing is suppressed because the dynamical temperature of the resolved particles is kept low. Fig.~\ref{fig:ae} illustrates the compositional mixing and migration of the resolved particles after $10^5$ yr. The color coding of the particles in Fig.~\ref{fig:ae} depicts the composition of the particle with respect to the starting composition at its current location. Green particles are composed of material predominantly from the particle's current location. Blue particles are composed of material predominately outside the particle's current location---from larger semi-major axis. Red particles are composed of material predominately inside the particle's current location---from smaller semi-major axis. Non-primary colors such as purple, cyan, or yellow indicate a mixture of the three categories listed above. The color of particles that contain material from various locations is determined by mass weighting the contributions from each primary color category: green---current location; blue---larger semi-major axis than the current location; red---smaller semi-major axis than current location. Using this coloring scheme, a noticable difference in the composition of planetesimals between the control simulations and those including full feedback from the debris becomes apparent by 100,000 yr. 

When dynamical friction from the debris is not included, the protoplanets seem to be evenly mixed. Consider the results from simulations 1a, 2a, and 3a (Fig.~\ref{fig:ae}). Most protoplanets are green, indicating that the majority of their composition is from their immediate starting surroundings. In each of these simulations there are also a few red and a few blue protoplanets, suggesting scattering of both the protoplanets and the smaller planetesimals that they have accreted has taken place. Simulation 4a also shows compositional mixing with three large non-primary colored protoplanets (one violet, one purple, and one pink). The colors of the protoplanets suggests that they are relatively heterogeneous in composition. In addition, there is no evidence of systematic migration of the protoplanets. When dynamical friction is included, however, there is little indication of compositional mixing because all particles have primary coloring. In simulations 1b, 2b, 3b, and 4b all particles are green or blue. However, there is strong indication of inward migration with simulations 2b, 3b, and 4b showing no particles at all at large semi-major axis and no red hued particles. The blue particles have moved inward from their original location. Most of the mass in these protoplanets originates from the protoplanet's original semi-major axis, which is larger than the protoplanet's current location.

Inward migration is an expected outcome from strong dynamical friction. 
There is still significant mass in debris at this time (Fig.~\ref{fig:debris}), thus the protoplanets are all experiencing dynamical friction from the background debris. As the protoplanets migrate inwards to smaller semi-major axis, due to circularization from dynamical friction and the accretion of debris, the protoplanets gravitationally scatter and stir each other because their gravitational spheres of influence ($10 - 15 R_H$) begin to overlap. This keeps eccentricities above zero and prevents dynamical friction from shutting off.

\begin{figure}
 \includegraphics[scale=0.4]{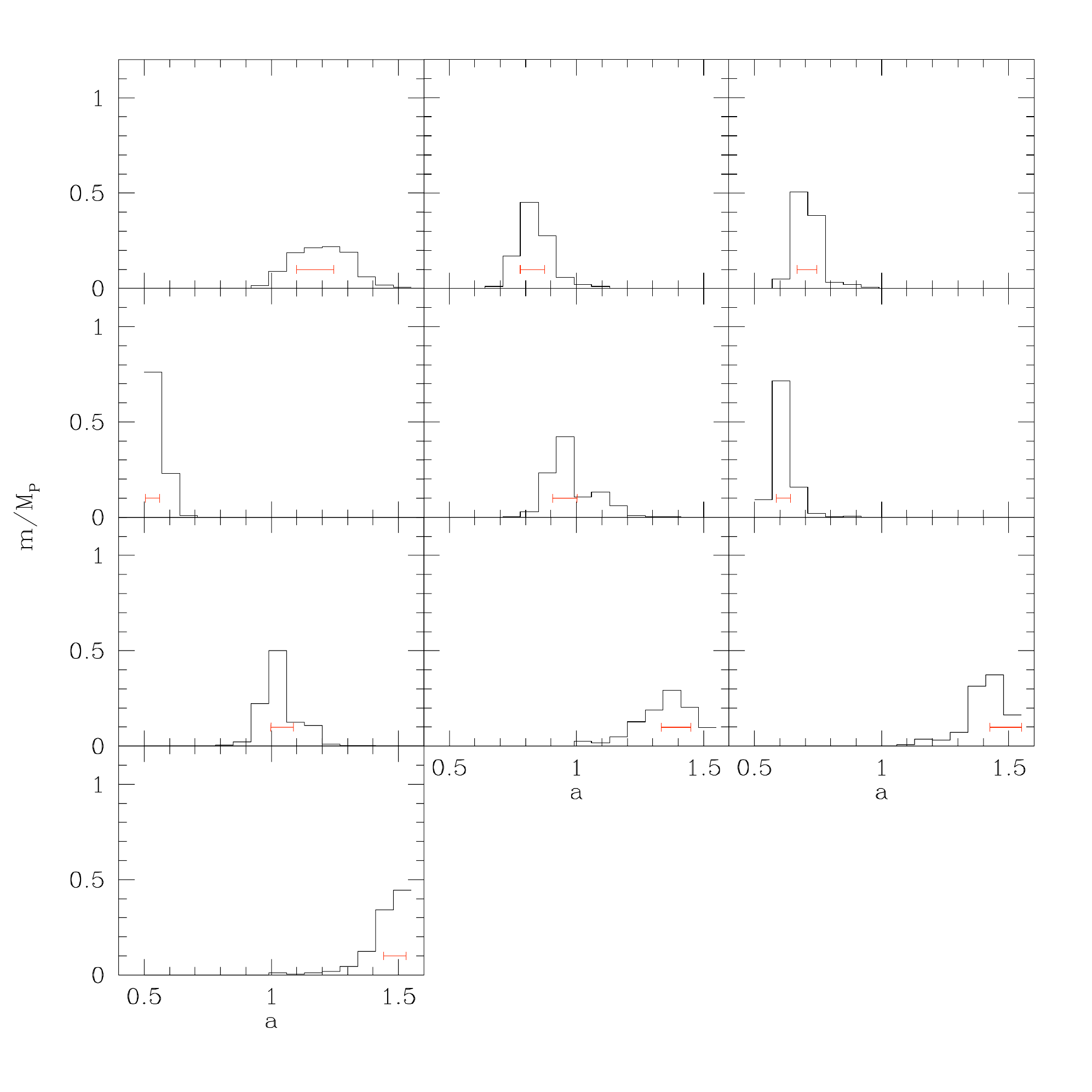}
 \caption{Composition histograms of the protoplanets---the most massive planetesimals---for simulation 3a after 100,000 yr.  The center of the red error bar indicates the current location of the protoplanet. The extent of the error bars (edge to edge) is $10 r_H$.
 \label{fig:ndyn}}
 \end{figure}

\begin{figure}
 \includegraphics[scale=0.4]{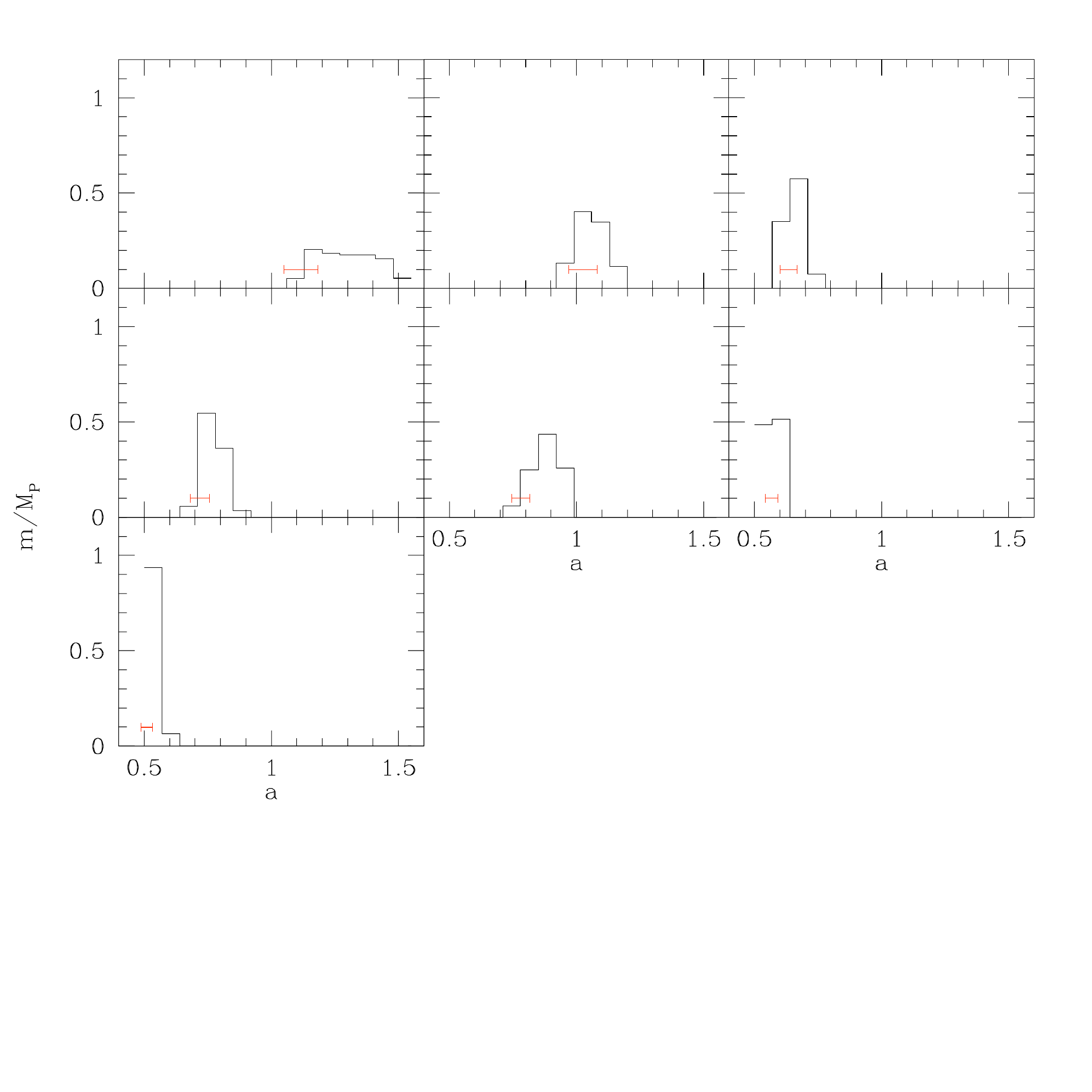}
 \caption{Same as Fig.~\ref{fig:ndyn} for simulation 3b, which includes dynamical friction from unresolved debris. In general, the protoplanets in this simulation are located at the left end of the range of their composition histograms, indicating inward migration.
 \label{fig:dyn}}
 \end{figure}

Figures \ref{fig:ndyn} and \ref{fig:dyn} show composition histograms of the most massive protoplanets after 100,000 yr for simulations 3a and 3b. The composition is binned in semi-major axis. The height of a given histogram bar represents the mass fraction of material from that semi-major axis bin in the protoplanet. The error bar denotes the current location of the protoplanet and the width of the error bar indicates $10 R_H$. The histograms are ordered in terms of mass of the protoplanet they describe, with the most massive in the upper left corner and least massive in the lower right. The protoplanets from simulation 3a are well mixed, containing, in general, roughly equal amounts of material originating both interior and exterior to their location (i.e. in most histograms, the error bar is in the middle). In the histograms from simulation 3b, on the other hand, the protoplanets tend to be on the left edge, which is consistent with inward migration. 

\subsection{Debris Evolution}\label{sec:debris} 

\begin{figure}
 \includegraphics[scale=0.4]{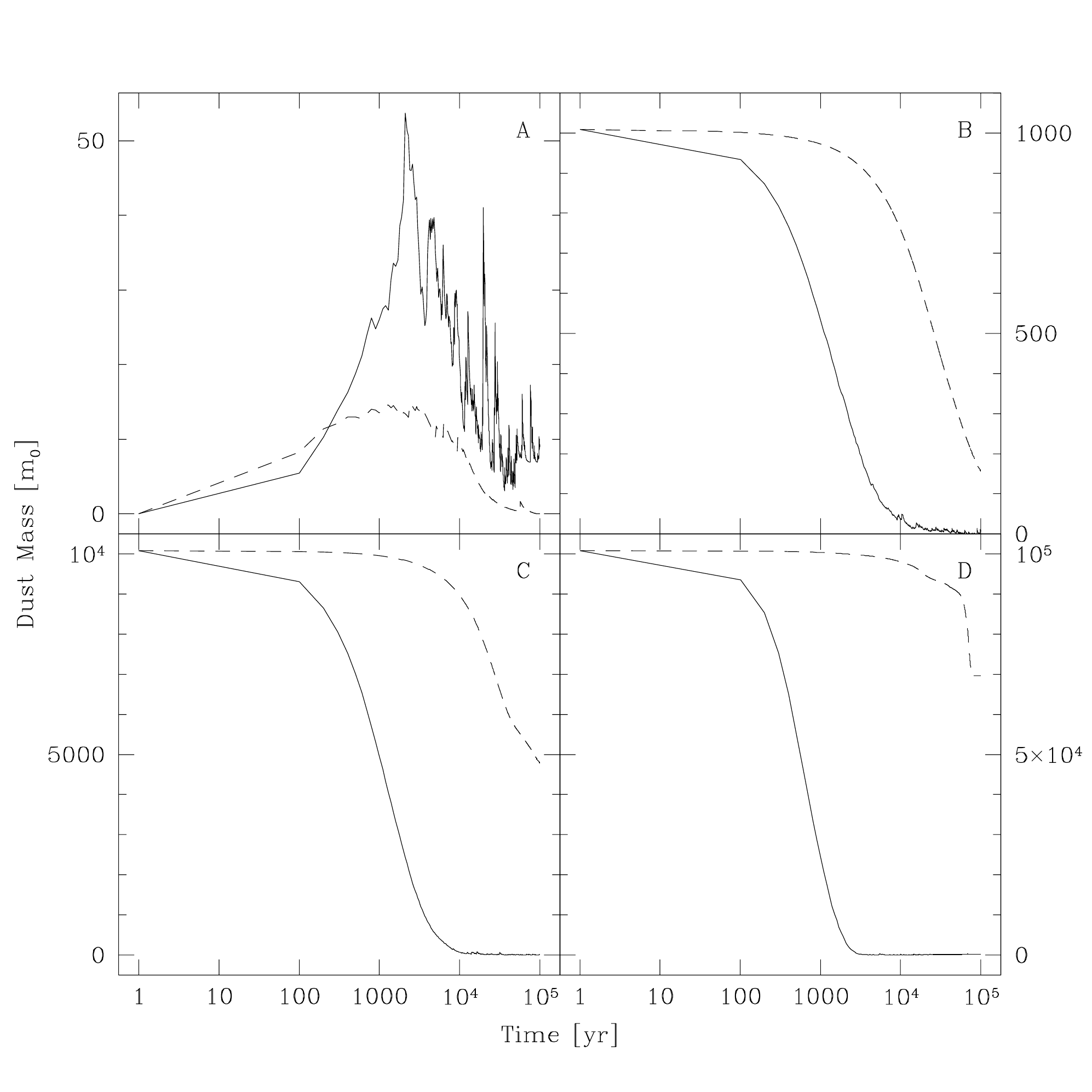}
 \caption{Total debris mass versus time. The solid lines are the control simulations 1a, 2a, 3a, and 4a (left to right, top to bottom). The dashed lines are simulations 1b, 2b, 3b, and 4b.
 \label{fig:debris}}
 \end{figure}

In general, the mass of unresolved debris decreases with time. The addition of dynamical friction from the unresolved debris slows the accretion of the debris onto the resolved particles. Figure \ref{fig:debris}A-D shows total mass of debris versus time for simulations 1a \& b, 2a \& b, 3a \& b, and 4a \& b, respectively. In general, the mass in background debris drops to less than half of the initial value by 1000 yr in simulations 2-4a. However, the decline in debris mass is much slower when dynamical friction is included. Dynamical friction from the background keeps the eccentricities of the planetesimals low, and as a result, this slows the accretion of the debris onto the planetesimals, meaning that the debris can have a lasting effect on the dynamics of the planetesimals. Recall that the mass accreted onto a given planetesimal \citep[Eq.~4 in][]{Leinhardt05} $\delta m = e \pi R^2 2 \pi a \rho \frac{\delta t}{P}$, where $e$ is the planetesimal eccentricity, $R$ is its radius, $a$ is the semi-major axis of its orbit, $\rho$ is the mass density of the debris in the annulus, $\delta t$ is the time since the last debris accretion update, and $P$ is the Keplerian period corresponding to $a$). The mass in the debris begins to drop eventually because the eccentricities of the protoplanets are small but non-zero and as a result dynamical friction from the unresolved debris causes inward migration. The protoplanets become crowded at small semi-major axis and gravitationally stir each other, which in turn increases their eccentricities. The accretion of the debris accelerates since the accretion is directly propotional to the eccentricity. 

The evolution of the debris in simulations 1a \& b differs from that seen in the simulations that begin with non-zero mass in debris. Surprisingly, even in this case, including dynamical friction of the background debris affects the evolution, though in a more subtle way than for simulations that begin with debris. In both simulations (1a \& 1b), the mass in debris increases (the result of debris-producing collisions) until a maximum is reached around 1000 yr, at which point the mass in debris begins to decrease. The larger planetesimals are now big enough that fewer and fewer collisions are disruptive. The maximum mass in debris reaches just over 50 and 10 $m_0$ for 1a and 1b, respectively. Even though the total mass in debris is 0.001 the total mass in planetesimals, it is enough to reduce the mass loss from a collision (see \S \ref{sec:collout} \& Fig.~\ref{fig:outcome}). 

  \begin{figure}
 \includegraphics[scale=0.43]{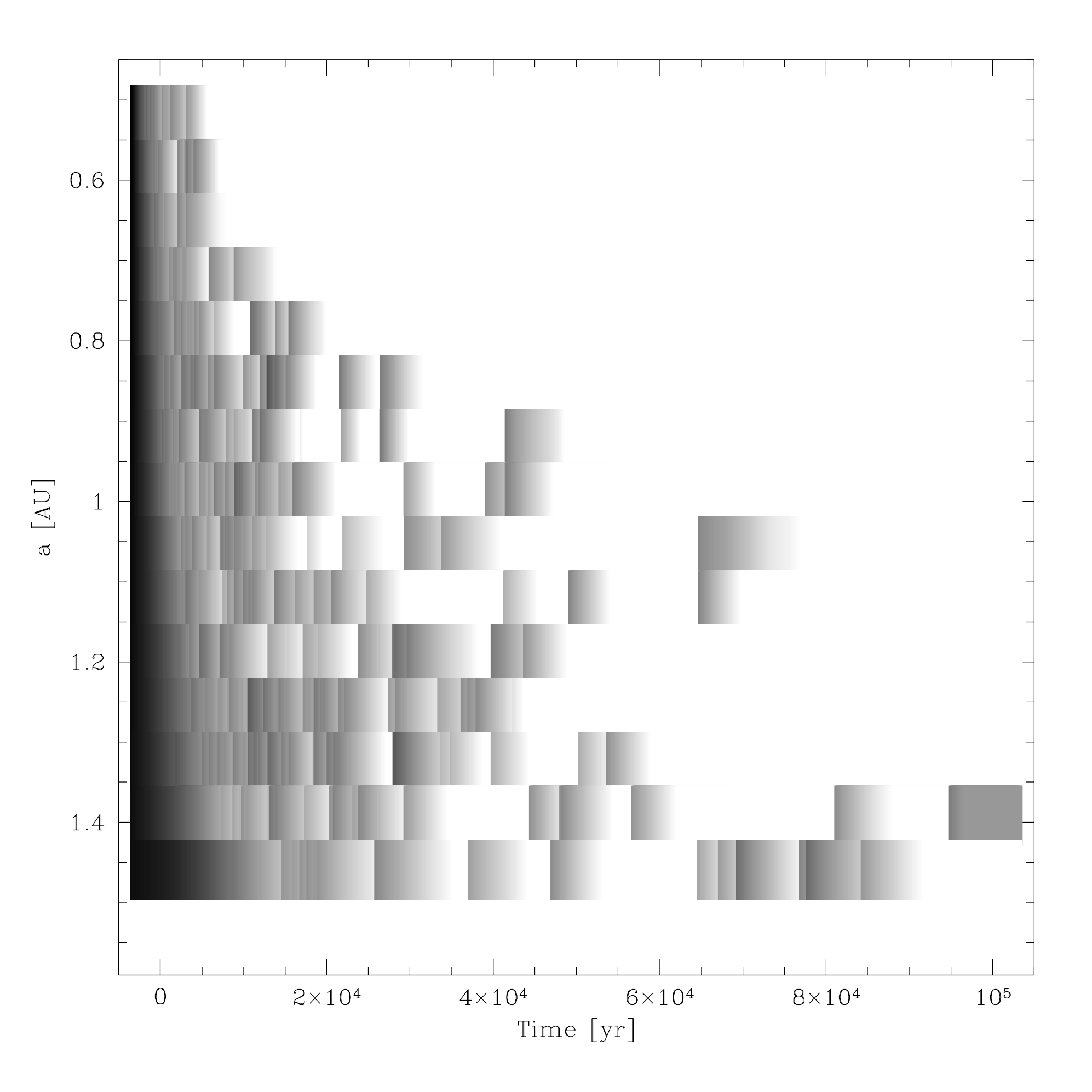}
 \caption{Debris image from simulation 3a: mass in unresolved debris as a function of time and semi-major axis. The mass in debris is shown in logarithmic grey scale from black to white---maximum to minimum (zero) mass. 
 \label{fig:imagenf}}
 \end{figure}
 
   \begin{figure}
 \includegraphics[scale=0.43]{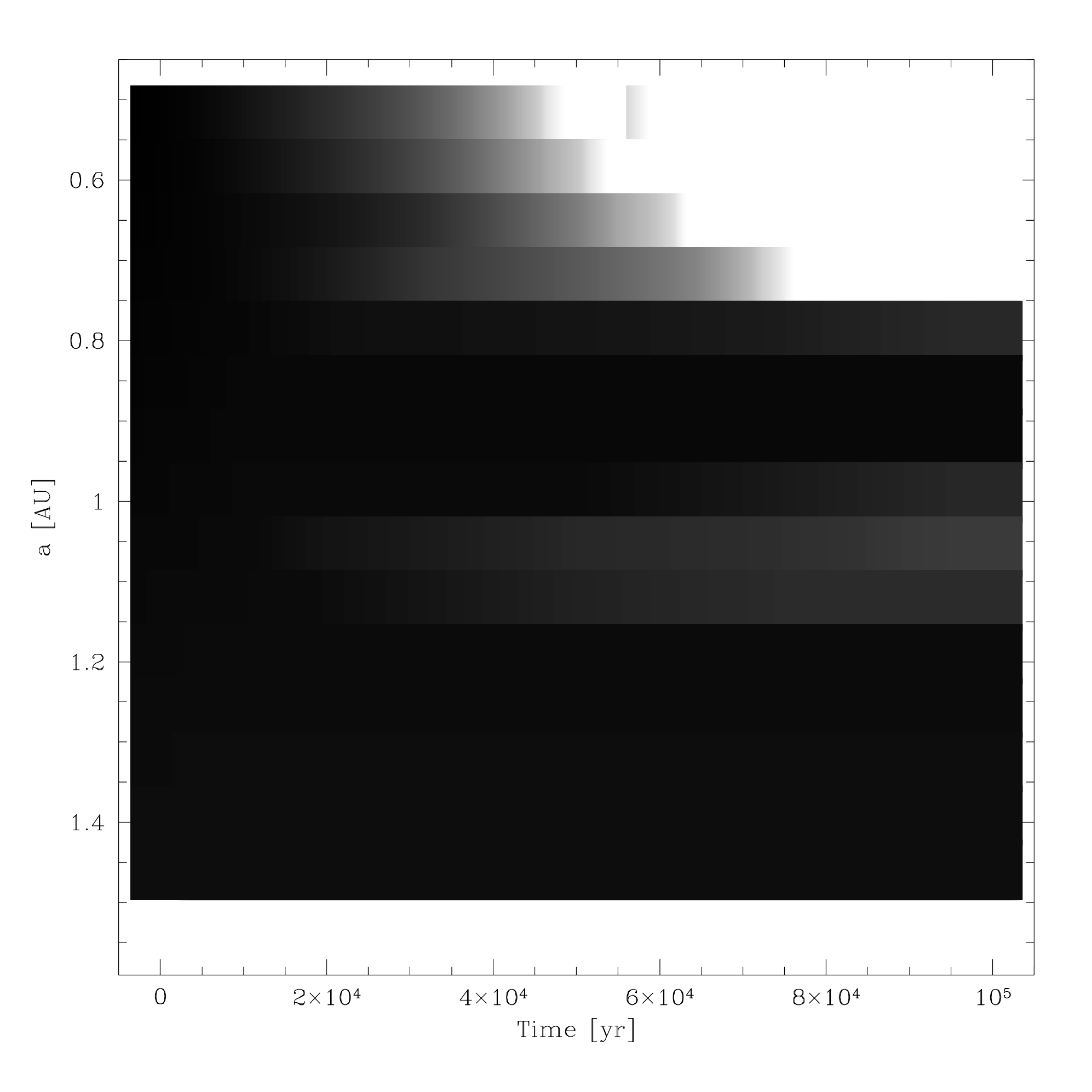}
 \caption{Same as Fig.~\ref{fig:imagenf} but for 3b.
 \label{fig:imagefr}}
 \end{figure}
 
\begin{figure}
 \includegraphics[scale=0.4]{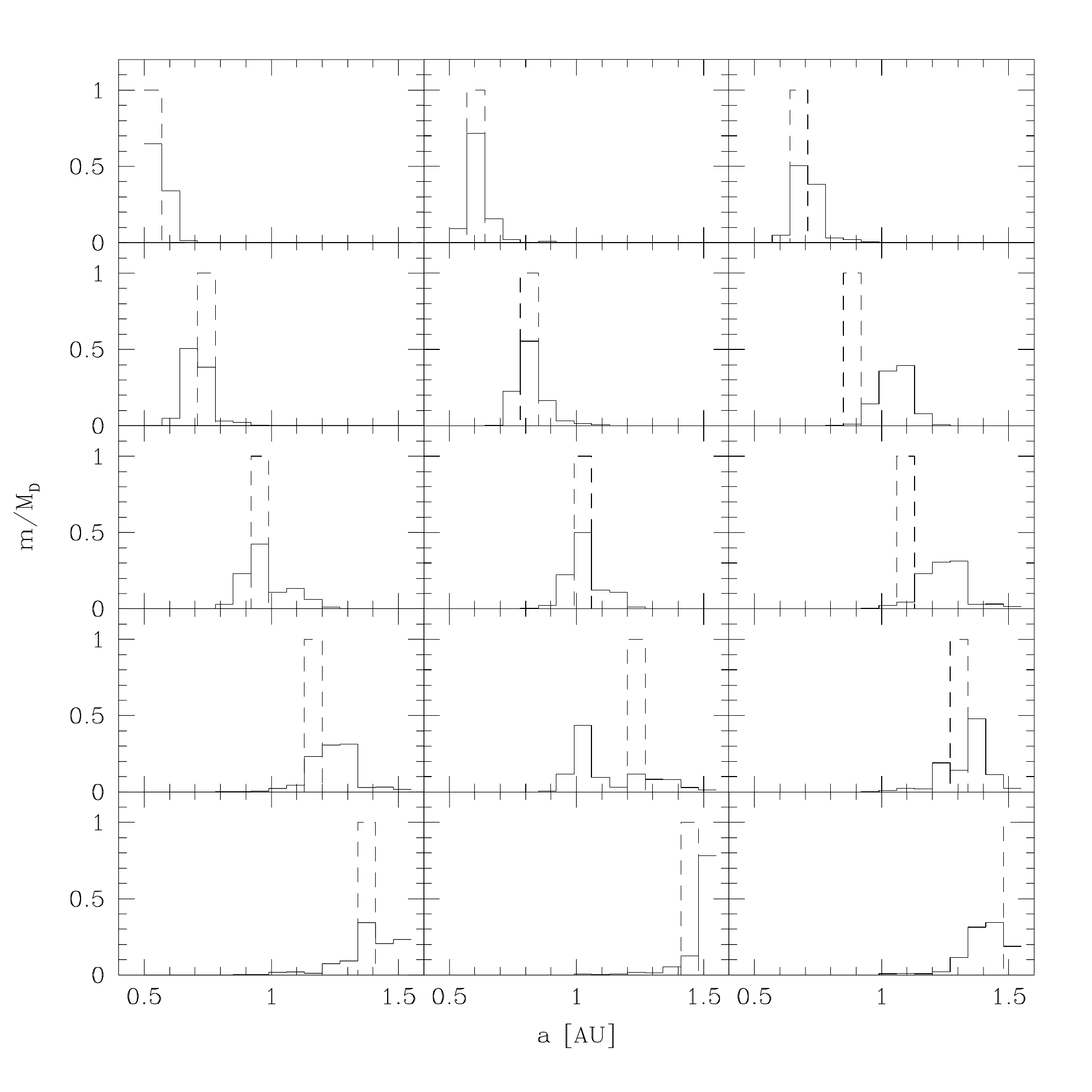}
 \caption{Composition histograms of the debris annuli from simulation 3a in solid and 3b in dashed after 100,000 yr. The semi-major axis of the debris annulus represented in each plot from upper left to lower right moves from the inner to the outer edge of the simulated annulus.  \label{fig:dhistnf}}
 \end{figure}

 In the control simulations the unresolved debris is effectively ``cleaned-up" at all semi-major axes, leaving effectively no debris anywhere by the end of the simulations at 100,000 yr. As an example, Fig.~\ref{fig:imagenf} shows the debris image of simulation 3a. The mass in unresolved debris is shown in grey-scale---black indicates the maximum mass, white indicates the minimum mass. The sharp transitions in the grey scale are the result of debris producing collisions. In contrast, the debris in the dynamical friction simulations is long lasting and is accreted onto protoplanets at late time and only at small semi-major axis. Figure \ref{fig:imagefr} shows the debris image for simulation 3b. The debris mass drops to zero only for the inner most four annuli. The effects of the debris even in the inner most annulus last until $\sim 4\times10^4$ yr. In addition, there are effectively no debris--producing collisions. This can be seen clearly by looking at the composition histograms of the debris annuli (Fig.~\ref{fig:dhistnf}). By 100,000 yr, the debris annuli in simulation 3a (solid lines in Fig.~\ref{fig:dhistnf}) are well mixed, indicating several debris-producing collisions. However, the composition histograms of simulations 3b (dashed lines in Fig.~\ref{fig:dhistnf}) show no mixing at all. All of the mass in a specified annulus is from that annulus. 
 
\begin{figure}
 \includegraphics[scale=0.4]{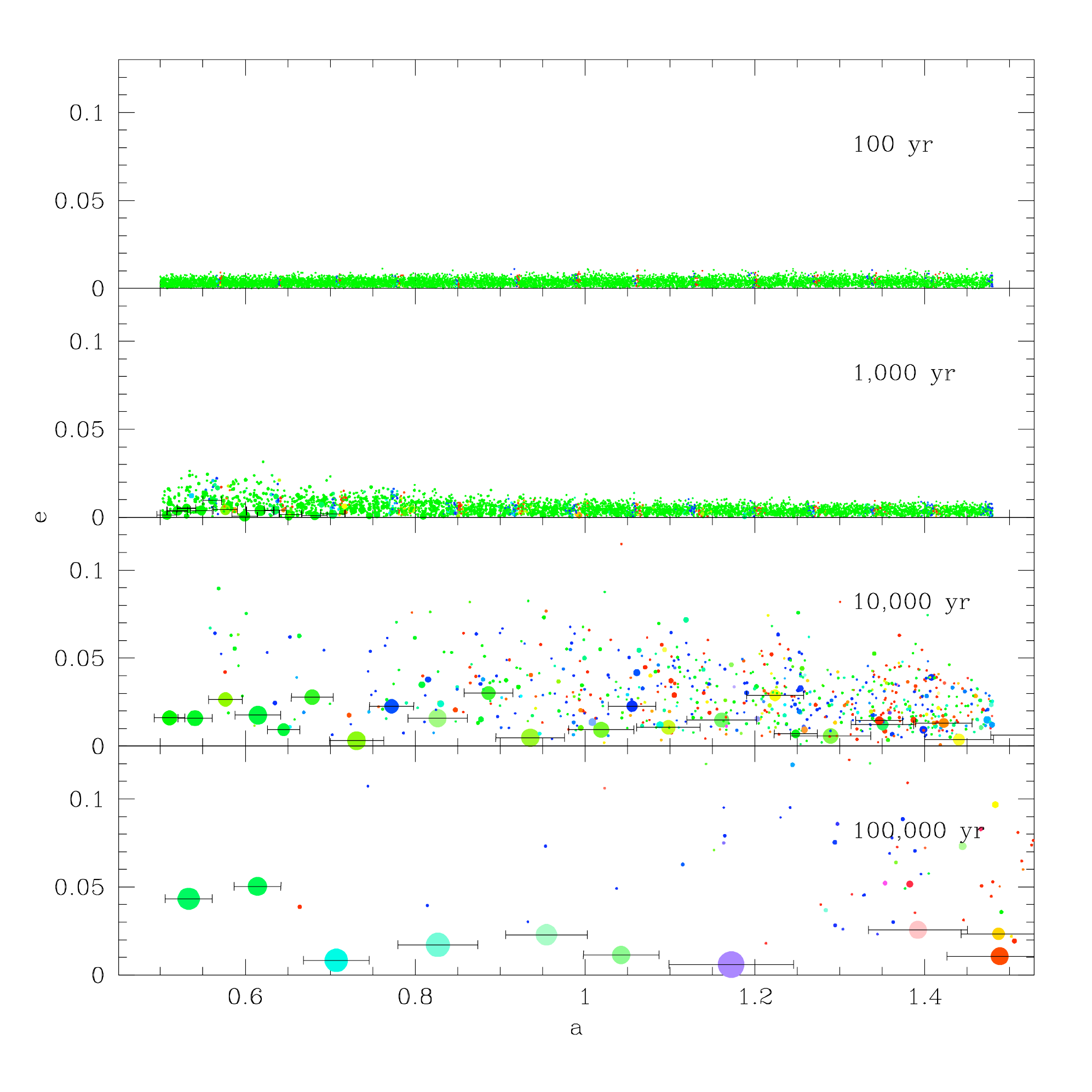}
 \caption{Snapshots of semi-major axis versus eccentricity for simulation 3a. The particle color coding denotes the amount of mixing. The size of the particle is proportional to its mass (see Fig.~\ref{fig:ae} caption). The particles with error bars (10 $r_H$) are at least two orders of magnitude more massive than the initial planetesimal mass. \label{fig:nfric10}}
 \end{figure}

\begin{figure}
 \includegraphics[scale=0.4]{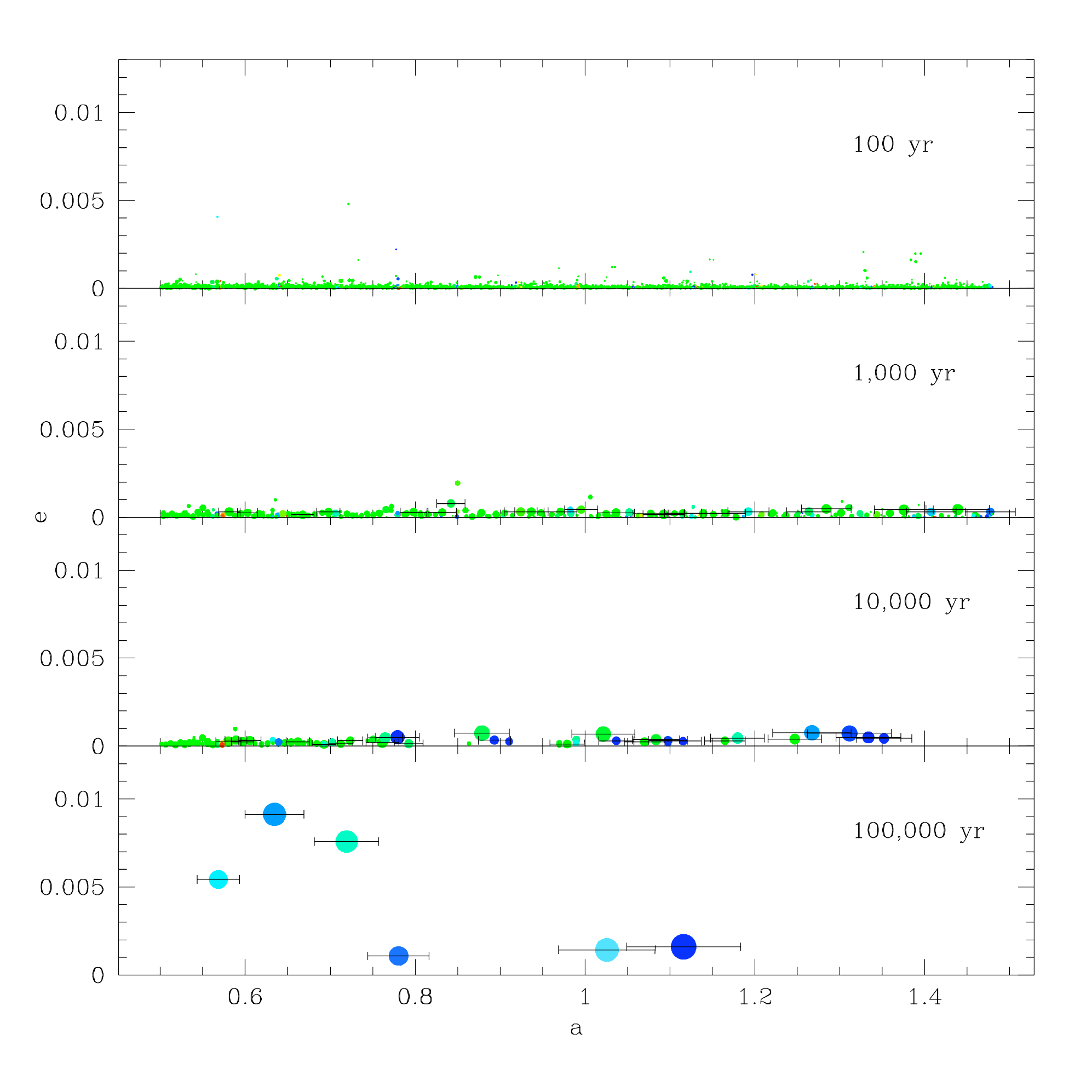}
 \caption{Same as Fig.~\ref{fig:nfric10} for simulation 3b, which includes dynamical friction from the unresolved debris. \label{fig:dynfric10}}
 \end{figure}

 Figures \ref{fig:nfric10} and \ref{fig:dynfric10} shows the time evolution of semi-major axis versus eccentricity for particles in simulation 3a and 3b, respectively. The eccentricity of the planetesimals in simulation 3a is large (0.07) by 10,000 yrs. In comparison, dynamical friction keeps the eccentricity of the planetesimals low in simulation 3b until there is significant inward migration. By 100,000 yrs, there are several protoplanets in 3b with significant eccentricity at $a < 0.8$. This is because inward migration has caused crowding and gravitational scattering at small semi-major axis, which has in turn resulted in efficient ``clean-up" of the debris.
 
\subsection{Growth Rate}\label{sec:growth}

\begin{figure}
 \includegraphics[scale=0.4]{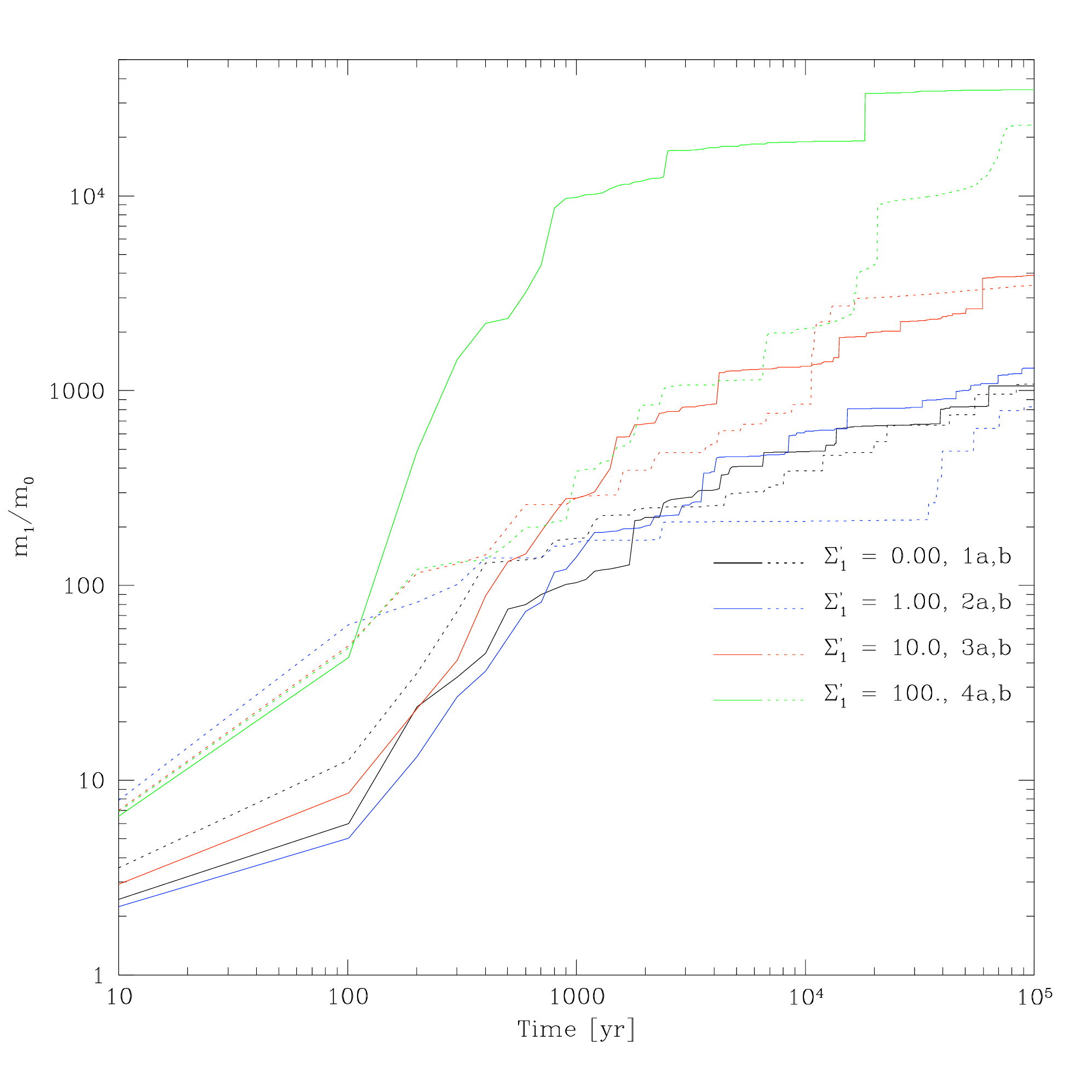}
 \caption{Mass of the instantaneous most massive particle as a function of time. Solid lines are from the control simulations (1a, 2a, 3a, 4a); dashed lines are results from simulations that contain feedback from the unresolved debris (1b, 2b, 3b, 4b).\label{fig:maxm}}
 \end{figure}

All of the control simulations go through two growth regimes: runaway growth followed by oligarchic growth (see solid lines in Fig.~\ref{fig:maxm}). In contrast, of the simulations that contain feedback from the unresolved debris, only the two with small amounts of initial debris (1b, 2b) go through two phases of growth (black and blue dashed lines, respectively). In runaway growth, the relative growth rate increases with mass ($
M^{-1}~ dM/dt \propto M^\alpha$, where $\alpha$ is a positive number), thus, more massive planetesimals grow more quickly than less massive planetesimals. In the oligarch phase, the power of the relative growth rate, $\alpha$, changes sign, so more massive protoplanets grow more slowly than less massive protoplanets, because the velocity dispersion of the planetesimals within several Hill spheres of the protoplanet is now dependent on the protoplanet's mass \citep{Kokubo98}. 

At early times, ($10^2-10^3$ yr), simulations 1-4a all have $\alpha > 0$, indicating onset of runaway growth. After $10^4$ yr, the power of the growth rate drops below zero, $\alpha \sim -0.6$, indicating a transition to oligarchic growth. In comparison, at early times, ($10^2-10^3$ yr) simulations 1 and 2b have a growth rate that is weakly positively dependent on $M$,  $\alpha \sim 0.1$. At later times ($t > 10^4$ yr), the relative growth rate becomes inversely dependent on mass, $\alpha \sim -0.5$.  Simulations 3b and 4b, which have more massive unresolved debris initial conditions, do not seem to go through a runaway growth regime. The relative growth rate is inversely proportional to $M$ even at early time, $\alpha \sim -0.4$. Both of these simulations (3b and 4b) have relatively constant power-law slopes for the entire $10^5$ yr. The growth rate of simulation 4b increases at very late time due to significant migration of the outer protoplanet. 

\subsection{Collision Outcome}\label{sec:collout}

\begin{figure}
 \includegraphics[scale=0.43]{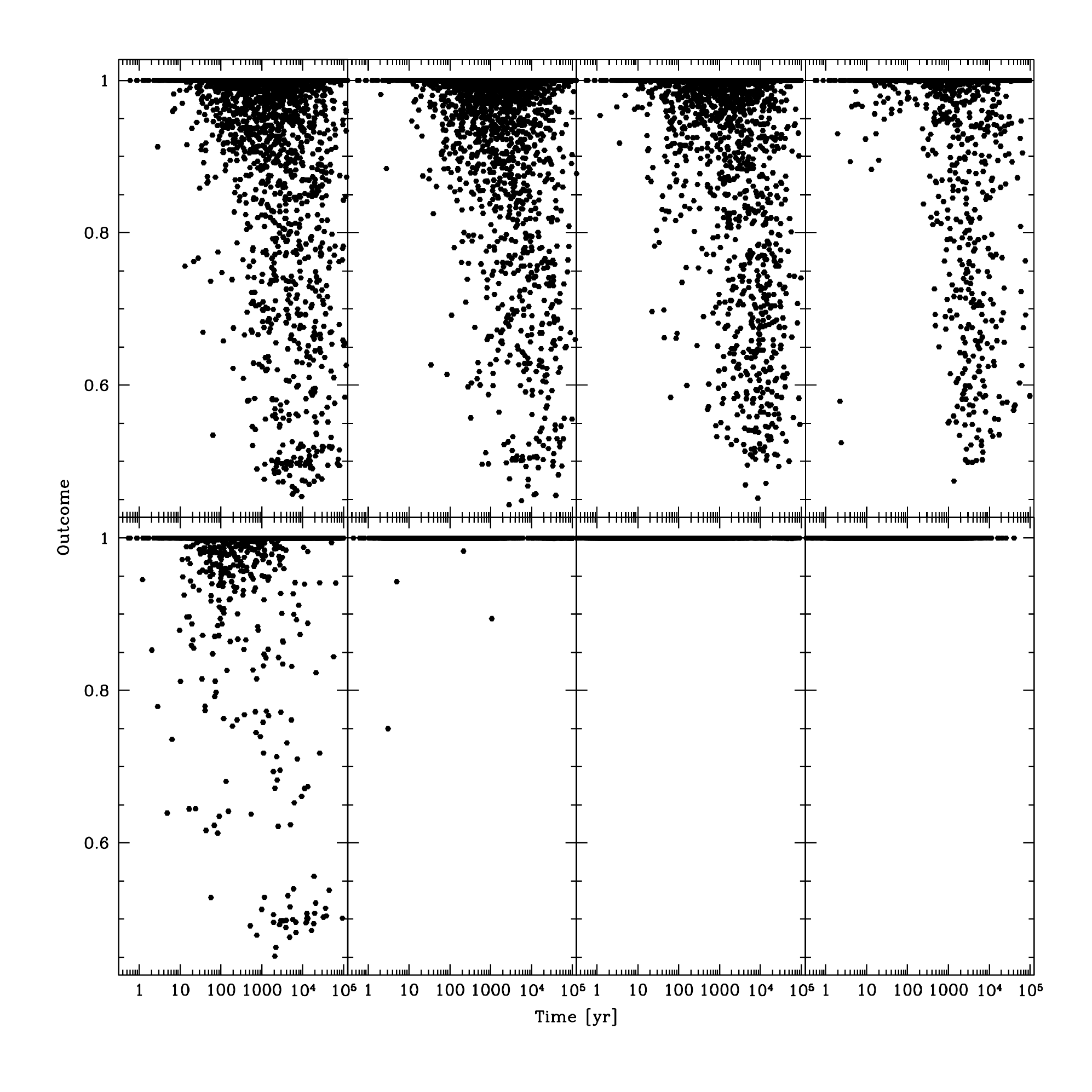}
 \caption{Collision outcome versus time in units of the total mass of the system for simulations 1a, 2a, 3a, and 4a (top row) and 1b, 2b, 3b, and 4b (bottom row). The outcome equals the mass of the largest post-collision remnant divided by the combined mass of the projectile and target. An outcome of 1 means that the projectile and target have merged. An outcome less than one means that some debris has been produced by the collision.
 \label{fig:outcome}}
 \end{figure}

The range of outcomes of planetesimal collisions differs dramatically between the control simulations and those that include dynamical friction (Fig.~\ref{fig:outcome}). In the control cases, there is a broad range of collision outcomes from disruptive debris-producing collisions to perfect merging events wherein all the mass from the projectile and target ends up in one post-collision remnant. In the control simulations, 6\% to 19\% of all collisions produce some debris compared to 0 to 4\% of all collisions in the simulations that include dynamical friction. When dynamical friction from the unresolved background is included, any significant background mass reduces the eccentricities of the planeteismals (Fig.~\ref{fig:ae}) and thus the impact speed of collisions. Therefore, the collision outcomes for simulations 2b, 3b, and 4b are effectively all perfect merging events. When there is no initial mass in debris (simulation 1b), the collisions are more violent, but the debris that is produced decreases the relative speeds of the impactors, reducing the number of debris-producing collisions compared to the control case (simulation 1a).

\begin{figure}
 \includegraphics[scale=0.43]{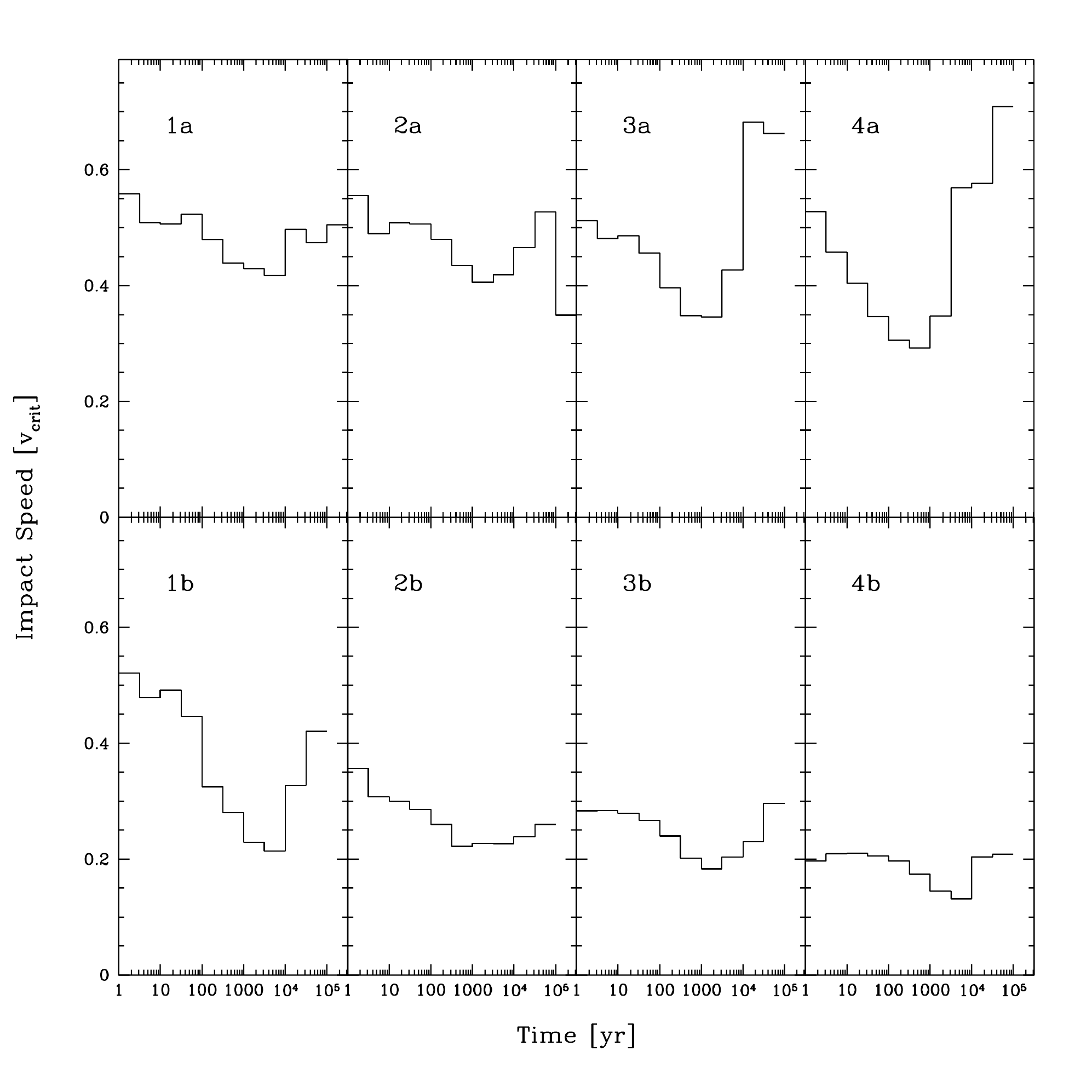}
 \caption{Histogram of average impact speed for all collision events between resolved planetesimals. The histogram bins are logarithmic in time.
 \label{fig:v}}
 \end{figure}

The trends in collision outcome are consistent with the mean impact speeds (Fig.~\ref{fig:v}). In the control case with no initial debris (sim.~1a) the mean impact speed stays relatively high during the entire simulation, fluctuating between 0.4 and 0.6 $v_{\rm crit}$, where $v_{\rm crit} = M \sqrt {\frac{6 G}{5 \mu R}}$ is the speed necessary to escape from an object with gravitational binding energy equal to the total mass of the colliding system, $M = M_{proj} + M_{targ}$ is the combined mass of the projectile and the target, and $R = (R_{proj}^3 + R_{targ}^3)^{1/3}$ is the radius of the combined mass assuming the target and the projectile have the same bulk density. When dynamical friction is included (Fig.~\ref{fig:v}, sim.~1b), the mean impact speed starts at about the same value but drops to about 0.2 $v_{\rm crit}$ by $10^4$ yrs. As a result of the slower impact speeds, there are fewer disruptive collisions in comparison to the control case. The trends for the other simulations are similar to those seen in simulations 1a and 1b. The control cases (Fig.~\ref{fig:v}, sim.~2a, 3a, \& 4a) have initial impact speeds around $0.5 v_{\rm crit}$ that drop with time, reaching a minimum around 1000 yr, at which point the mean impact speed begins to increase again, coming close to or surpassing the original mean impact speed by $10^5$ yr. The initial mean impact speeds for simulations 2b, 3b, and 4b are slower and the evolution of the impact speed is shallower than their control counter-parts. This explains why all of the collisions in these simulations result in perfect merging.

The total number of collisions is similar for all simulations, $\sim N$ over $10^5$ yr, where $N$ is the initial number of planetesimals in the simulation. If all collisions in a given simulation resulted in only one post-collision remnant, the number of collisions would equal $N-1$. Figure \ref{fig:outcome} shows that most collisions---even in the control cases---result in perfect merging and therefore guaranteed planetesimal growth.

\begin{figure*}
 \includegraphics[scale=0.4]{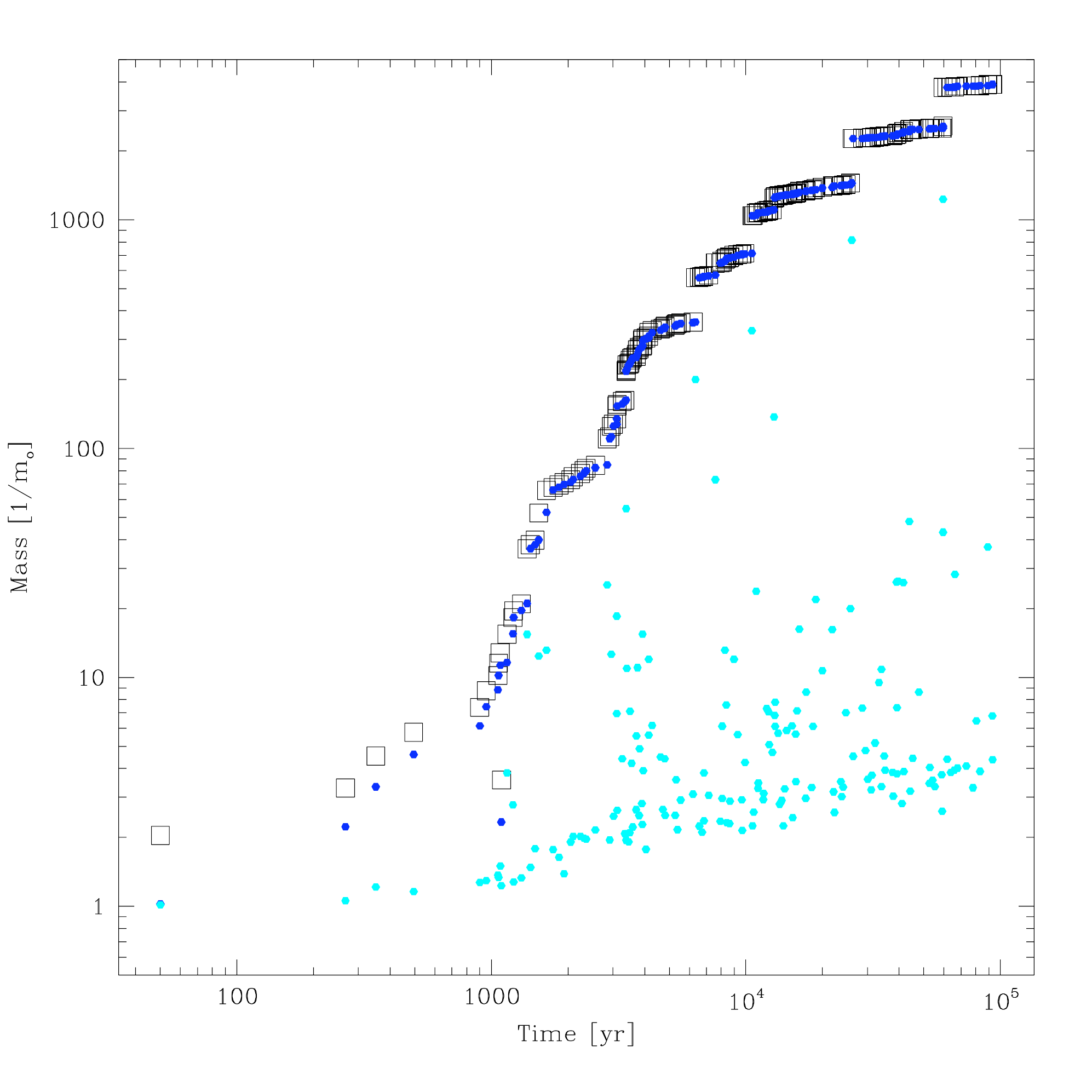} \includegraphics[scale=0.4]{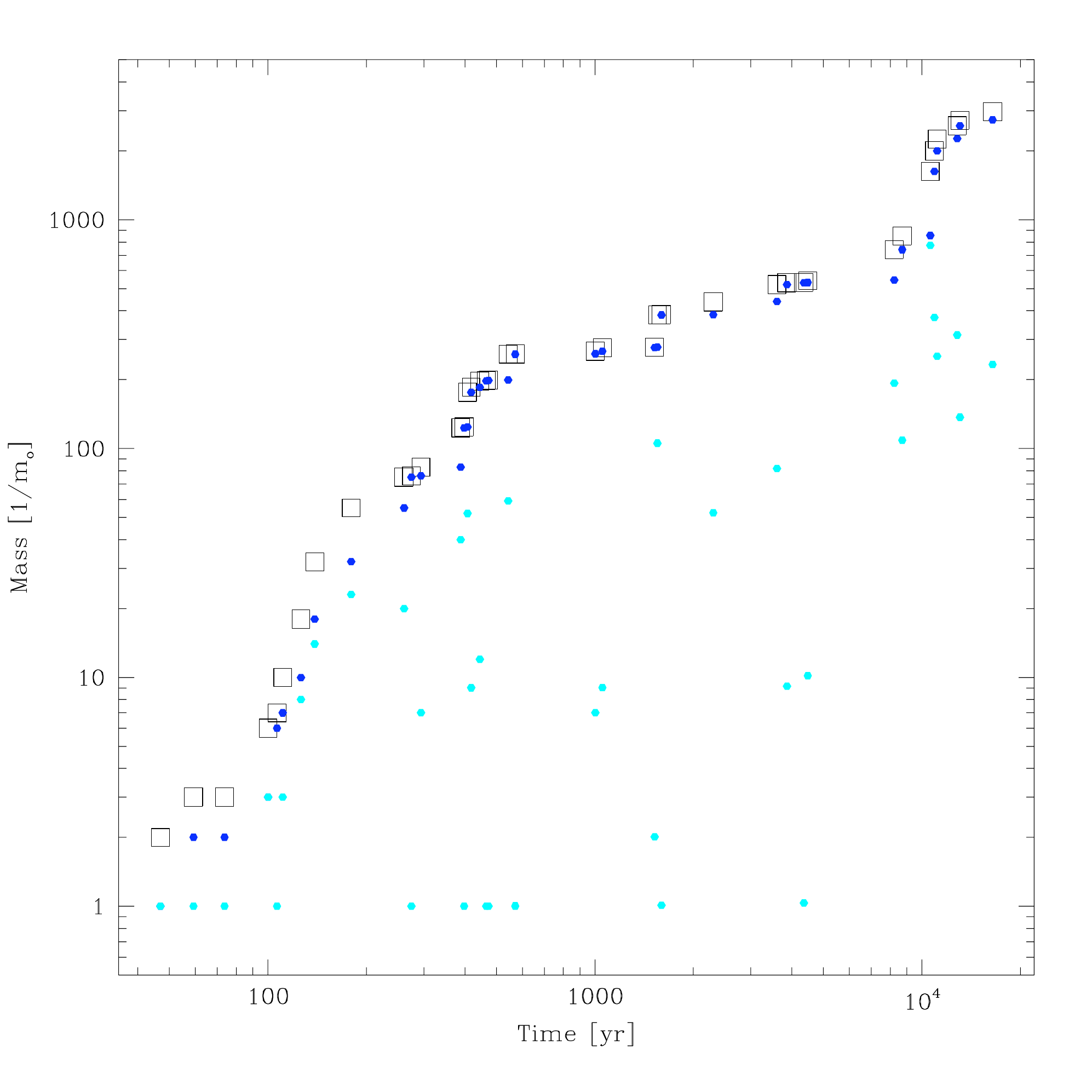}
 \caption{Mass evolution for one of the largest particles in simulation 3a (left) and 3b (right). Every collision involving the particle is depicted in these plots by a triplet of points: one blue point for the target, one cyan point for the projectile, and one open black square for the collision outcome.
 \label{fig:mergetree}}
 \end{figure*}
 
To highlight the difference in planetesimal evolution in both scenarios, Fig.~\ref{fig:mergetree} shows the collisional growth of one of the most massive particles in simulation 3a and 3b. (Fig.~\ref{fig:mergetree} is subtly different from Fig.~\ref{fig:maxm}, which shows the instantaneous most-massive particle and does not necessarily depict the growth of a single particle.) Every collision involving the chosen particle is indicated as a triplet of points: blue for the mass of the target, cyan for the mass of the projectile, and an open black square for the mass of the outcome. Effectively all collisions are perfect merging events in both simulations, meaning the outcome is a particle whose mass equals the mass of the projectile plus the mass of the target. In addition, the mass of the target in each collision involving the chosen particle  is the same as the outcome mass from the previous collision, therefore, there is little mass increase from accretion of unresolved debris. However, the particle in simulation 3a has at least 4 times as many collisions by $2\times 10^4$ yr. The majority of the collisions are with projectiles that are much smaller than the target. So although the particle in simulation 3a has many more collisions, the particle in simulation 3b has reached about the same mass by the end of the simulation.

\section{Discussion}\label{sec:disc}

The terrestrial planets in our solar system are co-planar and on low-eccentricity orbits. In general, it is argued that a damping force is required during the last phase of planet formation, when protoplanets grow into planets via catastrophic but infrequent collisions with other protoplanets, in order to produce a planetary system with low eccentricity and inclination. Without a damping mechanism, such as dynamical friction from a significant mass of small particles \citep{Obrien06,Goldreich04} or interaction with nebular gas \citep{Tanaka04}, the infrequent but strong gravitational encounters between protoplanets produce planets on highly eccentric, excited orbits. 

In this paper we have investigated the hypothesis that particle debris damps eccentricities, by conducting a series of numerical simulations of the middle phase of planet formation (during which planetesimals grow into protoplanets via planetesimal-planetesimal collisions) to determine if planetesimal collisions can produce and/or maintain the necessary massive background of small debris for the dynamical cooling of the large bodies in last phase of planet formation. Our simulations suggest that planetesimal collisions do not produce enough background material to provide significant dynamical friction during the last phase of planet formation. In this work we conclude that if the background material is responsible for the dynamical cooling of the protoplanets at late times, the material had to come from another source besides planetesimal-planetesimal collisions, or from collisions in an even earlier phase of planet formation.

It is possible that our simplified model for debris accretion is missing some important physics: for example our model does not include evolution of the unresolved debris, nor does it allow for dynamical heating of the background debris as a result of dynamical cooling of the large bodies. In addition, the model does not calculate the gravitational focusing of the debris by the protoplanets at late time. We model the debris semi-analytically; including heating, collisional evolution, or migration of the debris may decrease the efficiency of the dynamical friction, so the evolution of the protoplanets would take more time, but our general conclusions would not likely change. In fact we have already run some short simulations where the $V_{disp}$ was allowed to evolve with the velocity dispersion of the resolved planetesimals. The results are consistent with the previous statement. In the future we may consider increasing the realism of our collision model by including evolution of the debris to test this conclusion more rigorously.  

\section{Conclusions}

In this paper we have investigated the effect of initial background and impact-generated debris on the middle phase of terrestrial planet formation (planetesimals evolving into protoplanets). We have extended our numerical method from \citet{Leinhardt05} to include composition tracking and dynamical friction from unresolved debris. Numerical simulations using the extended method were compared to the results of numerical simulations using the original rubble-pile planetesimal collision model, which allows for erosion of planetesimals due to collisions but does not include dynamical cooling of the unresolved debris. We have found that with a mass of initial debris $\le 10\%$ of the mass in resolved planetesimals, the planetesimal evolution is qualitatively similar in both cases. If on the other hand there is significant initial debris mass ($\ge 10\%$ of the mass in resolved planetesimals) at the beginning of the simulation, the planetesimal growth modes change when dynamical friction of the debris is included. In particular, planetesimals grow concurrently when there is no background of smaller resolved planetesimals, which is the situation when the dynamical friction of the debris is not included. We find no situation in which there is enough debris produced from collisions to replenish the debris that is accreted onto the planetesimals. In addition, the composition of the resulting protoplanets is much more homogeneous than the protoplanets in the control simulations. When dynamical friction is included, there is less initial mixing but significant inward migration at later time. 

\section{Acknowledgements} ZML is currently supported by an STFC postdoctoral fellowship. DCR acknowledges support from NASA under grant no.\ NNX08AM39G. The simulations in this paper were carried out using the \texttt{borg} cluster administered by the Center for Theory and Computation in the Department of Astronomy at the University of Maryland. The authors thank K. Tsiganis (reviewer) and A. Young for comments that improved this manuscript.

\newpage
\bibliography{mn-jour,paper_dust}
\label{lastpage}
\end{document}